\documentclass[amsmath,12pt,amssymb,preprint,prd,aps,nofootinbib]{revtex4}
\usepackage{amsfonts} 
\usepackage{graphicx} 
\usepackage{epsfig}
\usepackage{multirow}
\usepackage{bm}

\begin{document}

\title{\textbf
{Production cross-sections and Radiative Decay widths \\
of Heavy Quarkonia in magnetized matter}}
\author{Amruta Mishra}
\email{amruta@physics.iitd.ac.in}
\author{Ankit Kumar}
\email{ankitchahal17795@gmail.com}
\affiliation{Department of Physics,
Indian Institute of Technology, Delhi, New Delhi - 110016}
\author{S.P. Misra}
\email{misrasibaprasad@gmail.com}
\affiliation{Institute of Physics, Bhubaneswar -- 751005, India}

\begin{abstract}
We study the production cross-sections and radiative decay widths 
of heavy quarkonia (charmonia and bottomonia) in magnetized nuclear matter.
The production cross-sections of the $\psi(3770)$
and $\Upsilon(4S)$, from the $D\bar D$ and $B\bar B$ scatterings
respectively, are studied from the medium modifications of the 
masses and partial decay widths to open charm (bottom) mesons,
of these heavy flavor mesons. Within a chiral effective model,
the masses of the vector and pseudoscalar charmonium (bottomonium) states
are calculated from the medium modification of a dilaton field, $\chi$,
which mimics the gluon condensates of QCD. The effects of
the Dirac sea (DS) and the anomalous magnetic moments (AMMs) 
of the nucleons are taken into consideration in the present study.
In the presence of a magnetic field,
there is mixing of the pseudoscalar (P) meson and the longitudinal 
component of the vector (V) meson (PV mixing), which leads
to appreciable modifications of their masses. 
The radiative decay widths of the vector (V) 
heavy quarkonia to the pseudoscalar (P) mesons
($J/\psi\rightarrow \eta_c(1S) \gamma$, $\psi(2S)\rightarrow 
\eta_c(2S) \gamma$ and $\psi(1D)\rightarrow \eta_c(2S) \gamma$
for the charm sector and $\Upsilon(NS)\rightarrow 
\eta_b(NS)\gamma$, $N$=1,2,3,4, for the bottom sector) 
in the magnetized asymmetric nuclear matter 
are also investigated in the present work.
The difference in the mass of the transverse component 
from the longitudinal component of the vector meson,
arising due to PV mixing, is observed as a double peak 
structure in the invariant mass spectrum of the
production cross-section of $\psi(3770)$.
This is observed to be appreciably more
pronounced as the value of the magnetic field is increased. 
For the bottomonium ($\Upsilon(4S)$) production cross-section, 
the effect of the PV mixing effect is observed to be marginal,
and there is observed to be a downward shift in the peak position 
due to the Dirac sea contributions. The modifications of the 
production cross-sections as well as the radiative decay widths
of the heavy quarkonia in the magnetized matter 
should have observable consequences 
on the production of these heavy flavour mesons
resulting from ultra-relativistic peripheral heavy ion collision 
experiments, where the created magnetic field can be extremely
large.
\end{abstract}
\maketitle
\section{Introduction}
The study of in-medium properties of the hadrons is an important 
topic of research in high energy physics, due to its relevance 
in relativistic heavy ion collision experiments \cite{Hosaka}. 
The modifications of the hadrons in the strongly interacting
matter resulting from the high energy nuclear collisions can affect 
the experimental observables. In the recent years, there have also
been a lot of studies on the effects of magnetic fields on 
the properties of the hadrons,
as the magnetic fields produced in peripheral ultra-relativistic 
heavy ion collision experiments, e.g., at LHC, CERN and RHIC, BNL 
are estimated to be huge \cite{tuchin}.
The strong magnetic fields created in the non-central 
heavy ion collisions decrease rapidly after the collison, 
as the ion remnants move away from the collision zone.
This leads to induced currents which slows down the decrease
of magnetic field by inducing a magnetic field in the
same direction as the external magnetic field, thereby
prolonging the lifetime of the magnetic field. 
The study of the time evolution of the magnetic field 
in the QCD plasma \cite{tuchin} is obtained from the solutions of the 
Maxwell's equations, and, assume Ohm's law, $\vec J=\sigma \vec E$.
However, the validity of Ohm's law has been questioned recently 
in Refs. \cite{Z_Wang,Igor_Particles_2022_5}, where it was
demonstrated that the relaxation time for the current, $\vec J$ to reach
the value of Ohm's law can be much larger than the lifetime of the
external electromagnetic field as well as the formation of
the QGP. This leads to what is termed as the 
`incomplete electromagnetic response' of the hot QCD matter
\cite{Z_Wang,Igor_Particles_2022_5}. Compared to the calculations
using Ohm's law \cite{Z_Wang,Igor_Particles_2022_5}, 
this was observed to lead to significant suppression
in the induced magnetic field (of the order of $10^{-2}$) 
for the non-expanding \cite{Z_Wang,Igor_Particles_2022_5}
as well as slowly expanding plasma \cite{Igor_Particles_2022_5}.
On the other hand, for rapidly expanding plasma, there was observed 
to be an enhancement of the electrical conductivity and hence
of the induced magnetic field \cite{Igor_Particles_2022_5}. 
The time evolution of the magnetic field depends
crucially on the electrical conductivity of the medium
and is still an open question. Further investigations
are needed on the topic, including obtaining a proper estimate 
of the electrical conductivity of the hot QCD matter.
However, since the heavy flavour mesons 
are created at the early stage when the magnetic field 
can still be large, the  magnetic field effects 
on the properties of the heavy flavour mesons could 
affect the experimental observables of these heavy ion collisions.

The heavy quarkonia (charmonia and bottomonia) are bound states of heavy
quark, $Q$ and heavy antiquark, $\bar Q$, $Q=c,b$). Studies of
these states in the presence of a gluonic field \cite{pes1,pes2,voloshin}
show that the modification of their masses is proportional to the change 
in the scalar gluon condensate in the leading order, 
treating the $Q$-$\bar Q$ separation to be small as compared 
to the gluonic fluctuations. Using the leading order formula, 
the in-medium masses of the vector charmonium states 
in nuclear matter have been studied from the medium change 
of the scalar gluon condensate, calculated 
in a linear density approximation in Ref.\cite{leeko}.
The study shows that the masses of the excited charmonium states 
have much larger shifts as compared to the ground state.
Using the QCD sum rule approach, the modifications
of the charmonium states arise due to the medium modifications
of the gluon condensate in the (hot) nuclear medium
\cite{kimlee,klingl,amarvjpsi_qsr,jpsi_etac_mag,moritalee}.
On the other hand, the modifications of the masses of the open 
heavy flavour mesons, which are bound states of a light ($q=u,d$) 
quark (antiquark) and a heavy (charm or bottom) antiquark (quark) 
arise due to the medium change of the light quark condensates 
in the nuclear medium within the framework of QCD sum rule approach
\cite{open_heavy_flavour_qsr,Wang_heavy_mesons,arvind_heavy_mesons_QSR}.
Within the quark meson coupling (QMC) model \cite{qmc}, 
the medium modifications of the masses of the open heavy 
flavour mesons in the nuclear medium
arise from the modifications of the scalar potentials of the constituent 
light quarks (antiquarks) of the open heavy flavour mesons
\cite{open_heavy_flavour_qmc}.

Within a chiral effective model \cite{Schechter,paper3,kristof1},
the broken scale invariance is incorporated in the model by introduction
of a scalar dilaton field, which simulates the gluon condensates of QCD.
The in-medium heavy quarkonium (charmonium and bottomonium) masses are
thus obtained from the medium changes of the scalar dilaton field
\cite{amarvdmesonTprc,amarvepja,AM_DP_upsilon} using the leading order
QCD formula \cite{pes1,pes2,voloshin,leeko}.
The in-medium masses of the open charm (bottom)
mesons ($D(\bar D)$, $B(\bar B)$, ${D_s}^\pm)$
within the chiral effective model
\cite{amdmeson,amarindamprc,amarvdmesonTprc,amarvepja,AM_DP_upsilon,DP_AM_Ds,DP_AM_bbar,DP_AM_Bs} arise from the interactions 
of these open heavy flavour mesons with the baryons and scalar mesons.
The partial decay widths of the heavy quarkonium states to 
the open heavy flavour mesons in (hot) strange hadronic matter
\cite{amarvepja}, arising from the 
mass modifications of these mesons have been studied
using a light quark creation model \cite{friman},
namely the $^3P_0$ model \cite{3p0_1,3p0_2,3p0_3,3p0_4} as well as
using a field theoretical model for composite hadrons
\cite{amspmwg,amspm_upsilon}.
Within the chiral effective model, 
in the presence of a magnetic field, 
the contributions of the Landau energy levels
to the number densities and the scalar densities of
the charged baryons, and, the lowest Landau level (LLL) contributions 
to the masses of the charged mesons are taken into account 
\cite{dmeson_mag,bmeson_mag,charmonium_mag,charmdecay_mag}.
In the presence of a magnetic field, there is mixing of the
pseudoscalar meson ($S=0$) and the longitudinal component 
of the vector meson ($S=1$) states (PV mixing) leading to appreciable 
modifications to their masses at high magnetic fields
\cite{charmonium_mag_QSR,charmonium_mag_lee,Suzuki_Lee_2017,Alford_Strickland_2013,Zhao_Prog_Part_Nucl_Phys_114_2020_103801,Quarkonia_B_Iwasaki_Oka_Suzuki}.
The heavy quarkonium
state in the presence of a constant magnetic field has been
studied as a non-relativistic two body system of a heavy quark
and heavy antiquark bound by a Cornell potential, using 
the Schrodinger equation 
in Refs. \cite{Alford_Strickland_2013,Zhao_Prog_Part_Nucl_Phys_114_2020_103801}.
In the presence of the magnetic field, the kinetic momentum 
is not a conserved quantity due to the breaking of the translational 
invariance, instead, the pseudomomentum, 
${\bf K}=\sum_{i=1}^2({\bf p}_i+q_i{\bf A}_i)$, is conserved 
\cite{Alford_Strickland_2013,Zhao_Prog_Part_Nucl_Phys_114_2020_103801}.
The studies show a strong dependence of the charmonia 
($\eta_c$ and longitudinal as well as transverse components
of $J/\psi$) energy eigenvalues (which are defined as `masses')
on the kinetic momentum ($\langle P_{kin}\rangle$) 
for strong magnetic fields 
\cite{Alford_Strickland_2013,Zhao_Prog_Part_Nucl_Phys_114_2020_103801}.
However, for the values of the magnetic field ($eB\le 10 m_\pi^2$) 
considered in the present study, the dependence on $\langle P_{kin}\rangle$ 
is observed to be rather marginal upto  $\langle P_{kin}\rangle \sim$ 2 GeV 
\cite{Zhao_Prog_Part_Nucl_Phys_114_2020_103801}.
In the present study of the production cross-sections and radiative
decay widths of the heavy quarkonia, 
the mixing of the pseudoscalar ($S=0$) and the vector meson ($S=1$) states 
for the charm sector are studied using a phenomenological interaction 
Lagrangian ($\sim {\tilde F}_{\mu \nu}(\partial ^\mu P) V^\nu$)
\cite{charmonium_mag_lee,Quarkonia_B_Iwasaki_Oka_Suzuki}, 
where the coupling parameter for the interaction is fitted
from the observed radiative decay width of $V\rightarrow P \gamma$. 
For zero spatial momenta of the pseudoscalar (P) and vector (V)
states, there is mixing of the pseudoscalar meson and the longitudinal
component of the vector meson (PV mixing), which leads to 
a drop (increase) in the mass of the pseudoscalar 
(longitudinal component of the vector meson) in the presence
of an external magnetic field, and the masses of the 
tranverse components of the vector mesons remain unaffected.
Due to lack of radiative data 
for the bottom sector, the PV mixing effect 
is considered using an interaction Hamiltonian
($\sim -{\mbox{\boldmath $\mu$}} \cdot {\bf B}$), which modifies
the masses of the pseudoscalar meson and the longitudinal component
of the vector meson due to their mixing
\cite{Alford_Strickland_2013,Zhao_Prog_Part_Nucl_Phys_114_2020_103801}.
These mass modifications due to PV mixings are considered 
for the study of the heavy quarkonia radiative decay widths
as well as the production cross-sections.
As we shall see later, the difference in the mass 
of the transverse component from the longitudinal component 
of the vector meson, arising due to PV mixing is observed 
as a double peak structure in the invariant mass spectrum of the
production cross-section of $\psi(3770)$, which is
observed to be significantly more pronounced as the strength 
of the magnetic field is increased. 
On the other hand, for the bottomonium ($\Upsilon(4S)$) 
production cross-section, it is observed that the PV mixing 
effects are marginal and  there is a downward shift in the peak position
due to the contributions of the magnetized Dirac sea.

Within the chiral effective model,
the effects of the Dirac sea (DS) of the nucleons
on the masses of the quarkonia and open heavy flavour mesons 
in magnetized nuclear matter have been studied in Refs.
\cite{Open_bottom_MC,Heavy_Quarkonia_masses_MC,charmdecay_mag_MC}
and are observed to be significant.
The effects of these mass modifications 
on the charmonium partial decay widths 
to $D\bar D$ mesons, have been studied using the $^3P_0$ model 
\cite{charmdecay_mag_MC},
and, the charmonium (bottomonium) decay widths to $D\bar D$ 
($B\bar B$) using the field theoretic model of composite hadrons 
\cite{charmdw_mag,open_charm_mag_AM_SPM,upslndw_mag,charmdw_mag_MC}.
The effects of PV mixing
as well as the Dirac sea (DS) effects of the nucleons
are taken into account for the study of the heavy quarkonia states.
The Dirac sea (DS) effects have been studied for the quark matter
using the Nambu-Jona-Lasinio (NJL) model \cite{Preis,menezes},
as well as, for nuclear matter within Walecka model
\cite{arghya,haber} as well as an extended linear sigma model
\cite{haber}. The effects of the Dirac sea of nucleons are taken into 
consideration by summing over the vector and scalar tadpole diagrams 
for obtaining the nucleon self-energy \cite{arghya}
within the chiral effective model 
\cite{Open_bottom_MC,Heavy_Quarkonia_masses_MC,charmdecay_mag_MC,charmdw_mag_MC}.
These are observed to lead
to an enhancement of the quark condenstaes (through the scalar fields)
with increase in the magnetic field, an effect called the `magnetic catalysis',
for zero density as well as for the baryon density, $\rho_B=\rho_0$, 
the nuclear matter saturation density, both for symmetric 
and asymmetric nuclear matter, when the AMMs are not taken 
into account. However, the opposite trend is observed 
for the quark condensates (inverse magnetic catalysis)
for $\rho_B=\rho_0$, when the AMMs of the nucleons 
are taken into account.
In peripheral ultra-relativistic heavy ion collisions,
the created matter has very small density.
In the present work, we study the effects of the magnetic field
on the production cross-sections 
of the heavy quarkonium states, $\psi(3770)$ and 
$\Upsilon(4S)$, which are the lowest states which decay to
the open charm (bottom) mesons in vacuum. In the magnetized nuclear
matter, these are studied
from the spectral functions of these states, obtained from 
their masses and the partial decay widths to $D\bar D$ ($B\bar B$),
taking into consideration the effects of PV mixing as well as
the Dirac sea of the nucleons
\cite{Heavy_Quarkonia_masses_MC,charmdw_mag_MC}. 
The in-medium decay widths of the heavy quarkonia are calculated
within a field theoretic model of composite hadrons \cite{charmdw_mag_MC}.
The effects of the in-medium masses of the heavy quarkonia 
on the radiative decay widths of $V\rightarrow P\gamma$
in the magnetized nuclear matter are also studied in the 
present work. The effects of the PV mixing as well as
Dirac sea on the heavy quarkonia production cross-sections as well as
radiative decay widths are observed to be significant, which 
should affect the experimental observables, e.g., the yields 
of the heavy quarkonia and the open heavy flavour mesons
produced in peripheral ultra-relativistic heavy ion collision
experiments.

The outline of the paper is as follows.
In section 2, we briefly recapitulate the chiral efective model
used for calculation of the masses of the heavy quarkonia 
(charmonia and bottomonia) as well as open charm (bottom) mesons
in magnetized asymmetric nuclear matter. 
The masses of these states are computed including the effects of
nucleon Dirac sea as well as the mixing
of the pseudoscalar (P) and vector (V) mesons (PV mixing). 
The in-medium decay widths of charmonia
(bottomonia) to open heavy flavour meson pairs $D\bar D
(B\bar B$), arising due to the mass modifications of the
initial and final mesons, are obtained using a field theoretical
model with composite hadrons.
Sections 3 describes the production cross-sections
of the heavy quarkonium states
(the charmonium state, $\psi(3770)$ and the upsilon state,
$\Upsilon (4S)$) arising from the scatterings
of $D\bar D$ and $B\bar B$ respectively.
The computation of the radiative decay widths of the 
vector (V) heavy quarkonia to  pseudoscalar (P) states 
in the magnetized nuclear matter is described in Section 4.
Section 5 describes the results of the present
investigation of the the production cross-sections 
and radiative decay wdths of heavy quarkonia
in magnetized nuclear matter. 
The summary of the findings of the present work
are presented in section 6.

\section{Masses and Decay widths of Heavy Quarkonium states}

\subsection{Masses}

The masses of the vector and pseudoscalar heavy quarkonium
(charmonium and bottomonium) states in the magnetized
nuclear matter are calculated within a chiral effective model.
The model is based on chiral symmetry and broken scale 
invariance of QCD, the latter incorporated into the model
through a logarithmic potential of a scalar dilaton field, $\chi$,
which mimics the gluon condensates of QCD.
Studies of the heavy quarkonium state in a gluon field, 
assuming  the distance between the heavy quark and antiquark
to be small as compared to the scale of gluonic 
fluctuations \cite{pes1,pes2,voloshin} leads to the
mass shift of the heavy quarkonium state to be proportional 
to the change in the gluon condensate in the medium,
in the leading order.
In the mean field approximation, the mass shifts of the heavy 
quarkonia in the magnetized nuclear matter are calculated 
within the chiral effective model
from the medium change of $\chi$
\cite{amarvdmesonTprc,amarvepja,AM_DP_upsilon},
which is related to the shift in the scalar gluon condensate.
The value of the dilaton field, $\chi$ is obtained from
the solution of the coupled equations of motion of $\chi$ and the scalar 
(non-strange isoscalar, $\sigma$, strange isoscalar, $\zeta$
and non-strange isovector $\delta$) fields, for given values of 
the baryon density, $\rho_B$, the isospin asymmetry parameter,
$\eta=(\rho_n-\rho_p)/(2\rho_B)$ (with $\rho_n$ and $\rho_p$
as the neutron and proton number densities), and, the 
magnetic field, $B$ (chosen to be along z-direction).
The anomalous magnetic moments (AMMs) of the nucleons
\cite{dmeson_mag,bmeson_mag,charmonium_mag}, and, the
contributions due to the nucleon Dirac sea, are taken
into account to study the masses of the 
charmonium ($J/\psi$, $\psi'\equiv \psi(2S)$, $\psi''\equiv \psi(1D)$, 
$\eta_c\equiv \eta_c(1S)$ and $\eta_c'\equiv \eta_c(2S)$) 
and the bottomonium ($\Upsilon(1S)$, 
$\Upsilon(2S)$, $\Upsilon(3S)$, $\Upsilon(4S)$, $\eta_b\equiv \eta_b(1S)$,
$\eta_b'\equiv \eta_b(2S)$, $\eta_b(3S)$ and $\eta_b(4S)$)
states in magnetized (nuclear) matter. 
These masses have additonal contributions due to
the mixing of pseudoscalar (P) meson and the
longitudinal component of the vector (V) meson
(PV mixing) in the presence of a magnetic field
\cite{charmonium_mag_QSR,charmonium_mag_lee,Suzuki_Lee_2017,Alford_Strickland_2013,Zhao_Prog_Part_Nucl_Phys_114_2020_103801,Quarkonia_B_Iwasaki_Oka_Suzuki,charmdw_mag,open_charm_mag_AM_SPM}.
These PV mixing effects ($J/\psi-\eta_c$, $\psi(2S)-\eta_c (2S)$, $\psi(1D)-\eta_c (2S)$
for the charmonium states and $\Upsilon (1S)-\eta_b$, 
$\Upsilon (2S)-\eta_b(2S)$, $\Upsilon (3S)-\eta_b(3S)$,
$\Upsilon (4S)-\eta_b(4S)$ for the bottomonium states),
are considered in the present study 
of the  production cross-sections  and radiative decays
of heavy quarkonia.

\begin{figure}
\vskip -2.2in
    \includegraphics[width=0.9\textwidth]{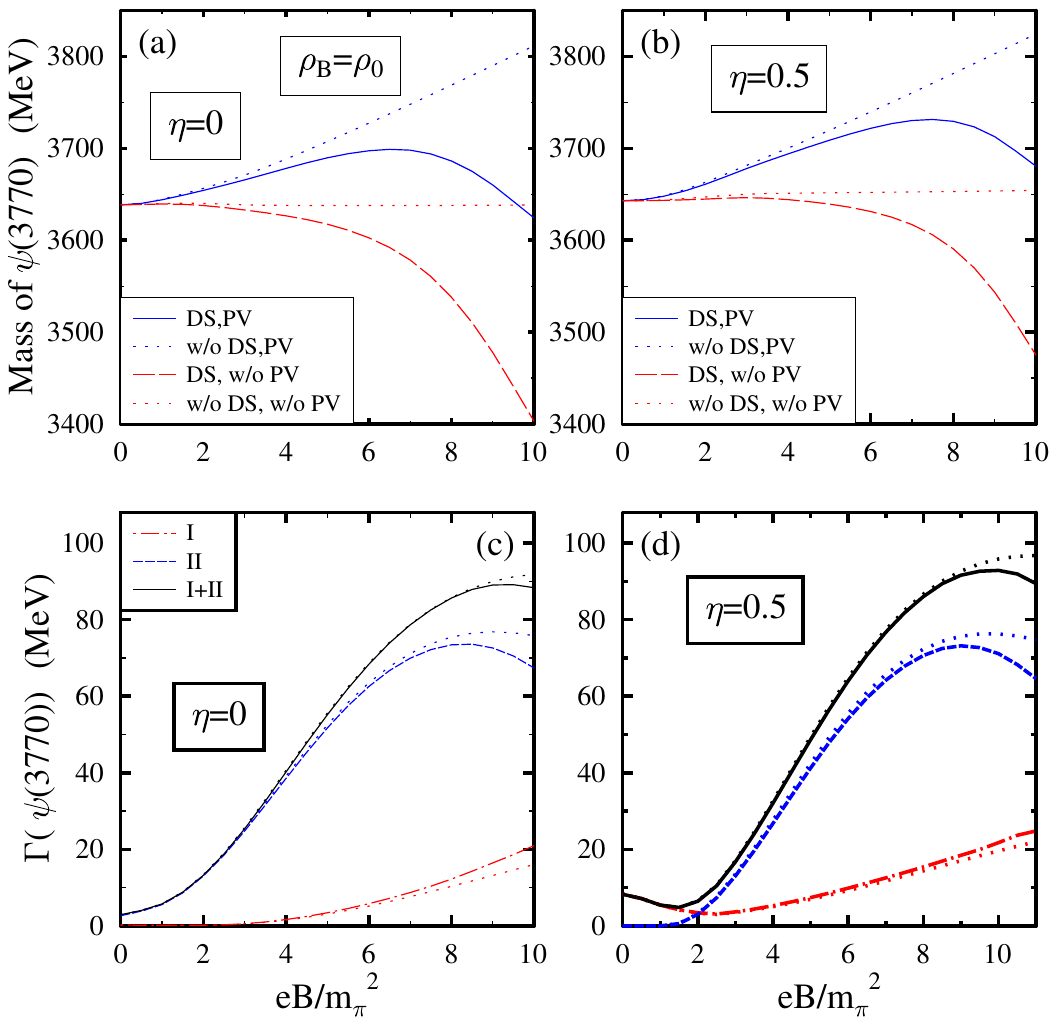}\hfill
\vskip -0.3in
    \caption{The effects of Dirac sea (DS) and PV mixing 
on  the masses and decay widths of $\psi(3770)$ 
are shown as functions of $eB/{m_\pi^2}$,
for the symmetric ($\eta$=0) and asymmetric 
(with $\eta$=0.5) nuclear matter.    
The decay width of $\psi(3770)$ to (I) $D^+D^-$, (II) $D^0 \bar {D^0}$
and (III) $D^+D^-$ as well as $D^0 \bar {D^0}$, are shown in panels
(c) and (d) with Dirac sea effects and compared with the results
without Dirac sea effects (shown as dotted lines).}
\label{Mass_DW_3770_rhb0}
\end{figure}

\begin{figure}
\vskip -2.2in
    \includegraphics[width=0.9\textwidth]{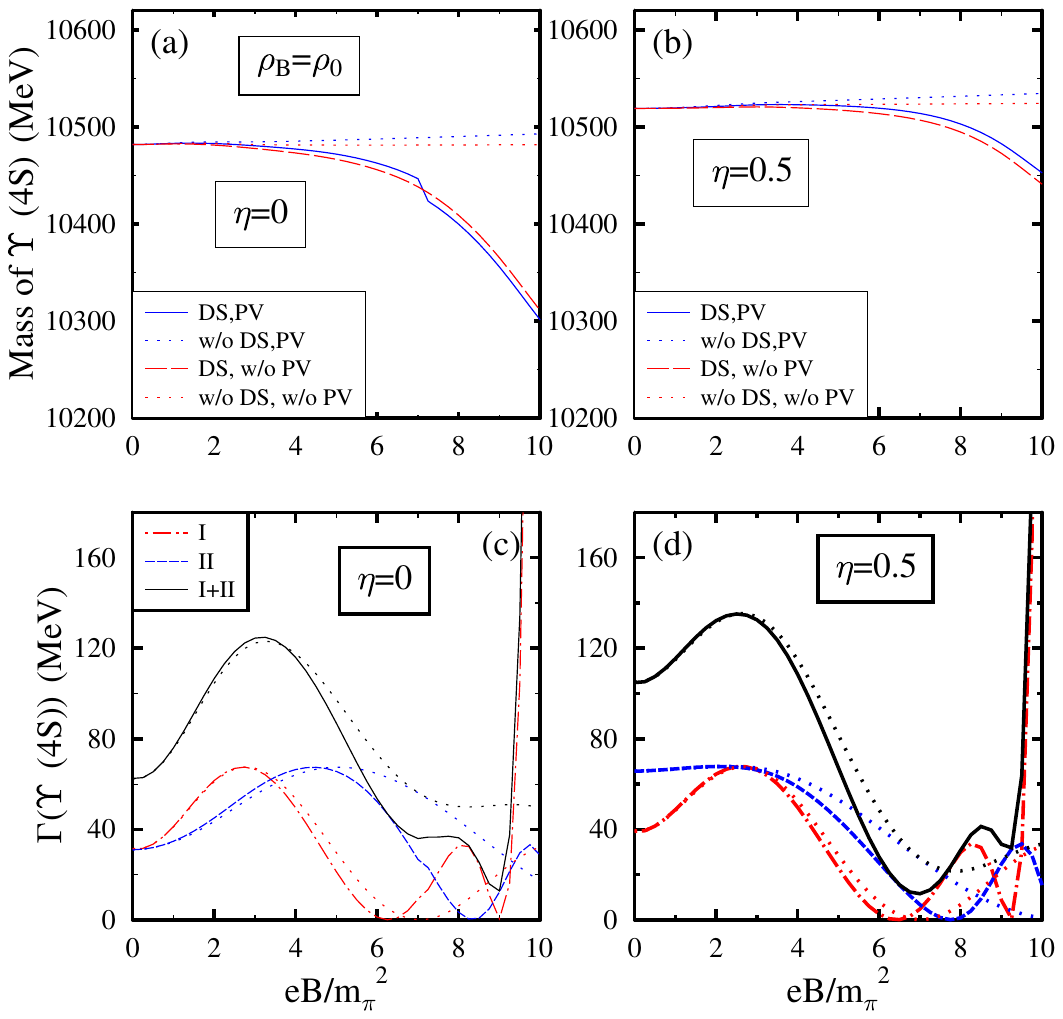}\hfill
\vskip -0.3in
    \caption{The effects of Dirac sea (DS) and PV mixing 
on  the masses and decay widths of $\Upsilon(4S)$ 
are shown as functions of $eB/{m_\pi^2}$,
for the symmetric ($\eta$=0) and asymmetric 
(with $\eta$=0.5) nuclear matter.    
The decay widths of $\Upsilon(4S)$ to (I) $B^+B^-$, (II) $B^0 \bar {B^0}$
and (III) $B^+B^-$ as well as $B^0 \bar {B^0}$, are shown 
including the Dirac sea effects and compared with the results
without Dirac sea effects (shown as dotted lines).}
\label{Mass_DW_Upsilon4S_rhb0}
\end{figure}

\begin{figure}
\vskip -2.2in
    \includegraphics[width=0.9\textwidth]{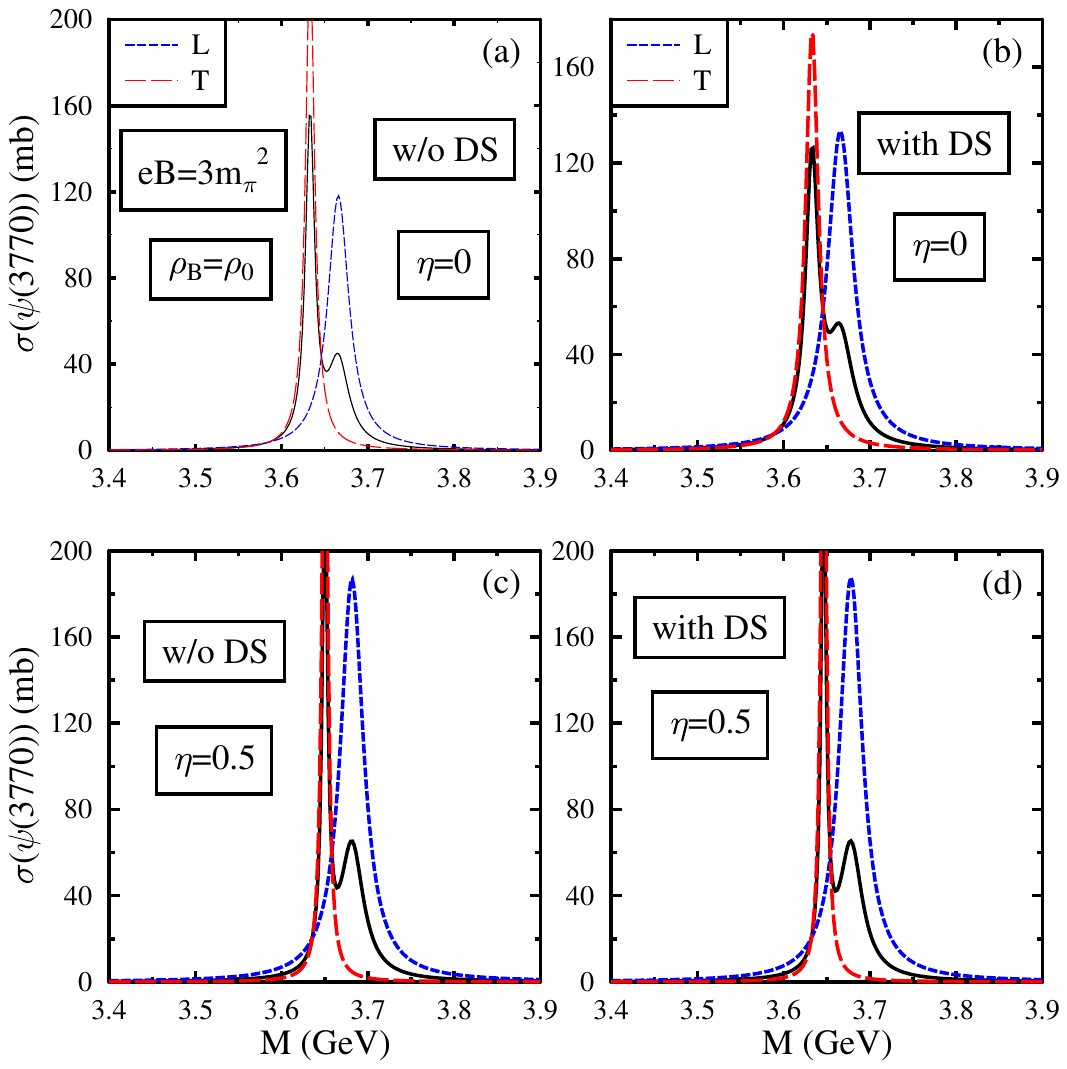}\hfill
\vskip -0.3in
    \caption{Production cross-section of $\psi(3770)$ due to
scattering of $D\bar D$ mesons for $eB=3 m_\pi^2$, showing 
the contributions from the longitudinal (L) and transverse 
(T) components of $\psi(3770)$.}
    \label{Cross_section_psi3770_rhb0_3mpi2}
\end{figure}

\begin{figure}
\vskip -2.2in
    \includegraphics[width=0.9\textwidth]{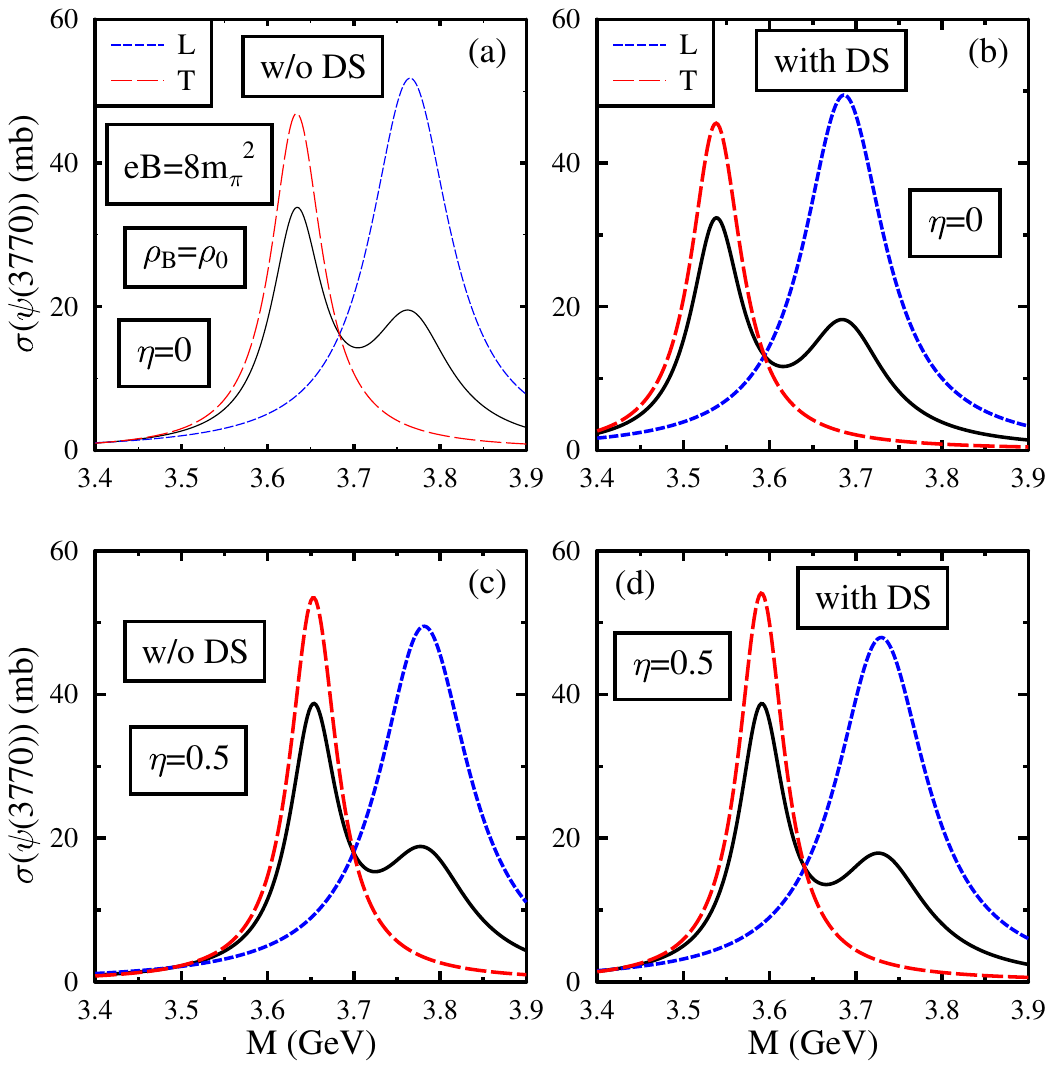}\hfill
\vskip -0.3in
    \caption{Same as figure \ref{Cross_section_psi3770_rhb0_3mpi2},
for $eB=8 m_\pi^2$.}
    \label{Cross_section_psi3770_rhb0_8mpi2}
\end{figure}

\begin{figure}
\vskip -2.2in
    \includegraphics[width=0.9\textwidth]{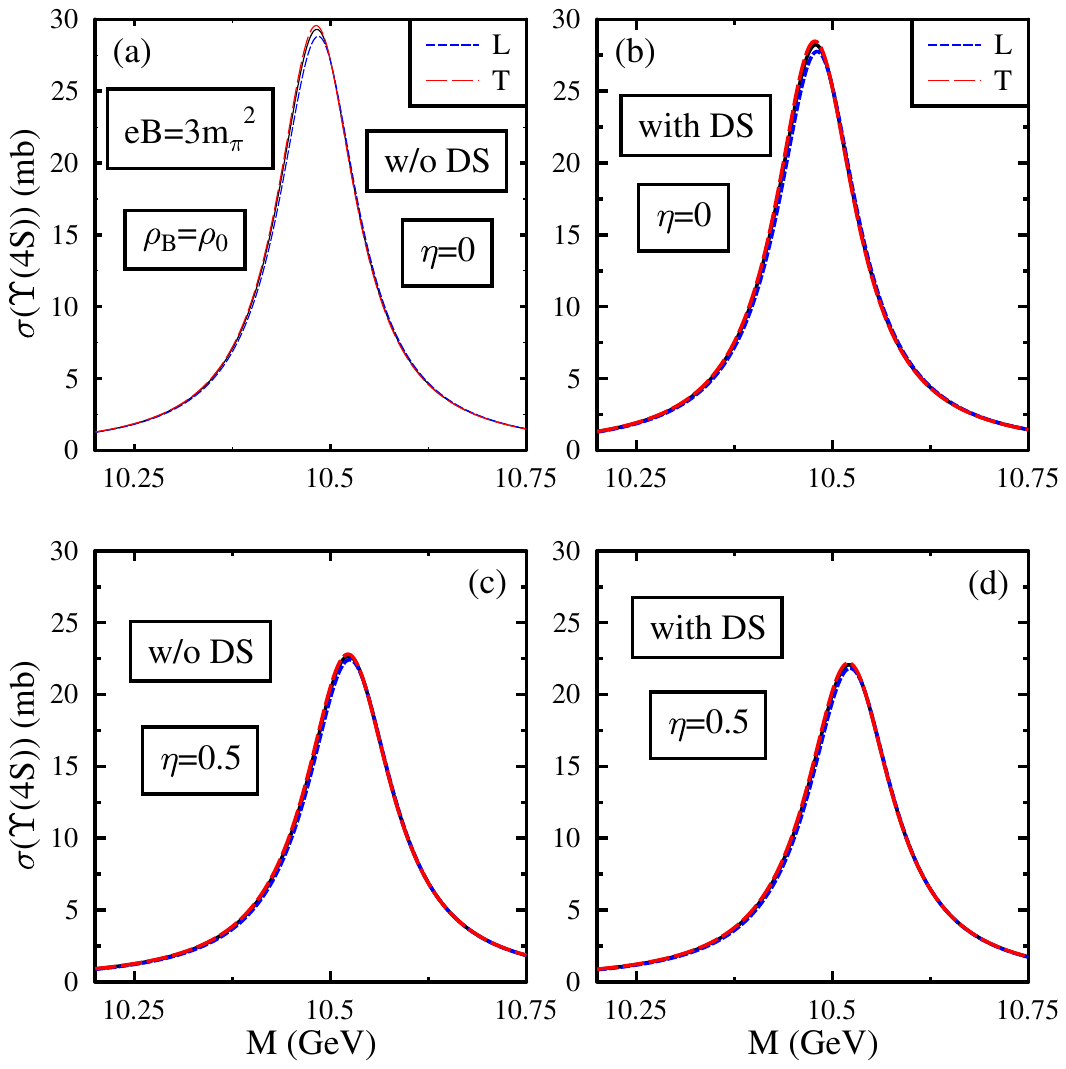}\hfill
\vskip -0.3in
    \caption{Production cross-section of $\Upsilon(4S)$ due to
scattering of $B\bar B$ mesons for $eB=3 m_\pi^2$, showing 
the contributions from the longitudinal (L) and transverse 
(T) components.}
    \label{Cross_section_Upsilon4s_rhb0_3mpi2}
\end{figure}

\begin{figure}
\vskip -2.2in
    \includegraphics[width=0.9 \textwidth]{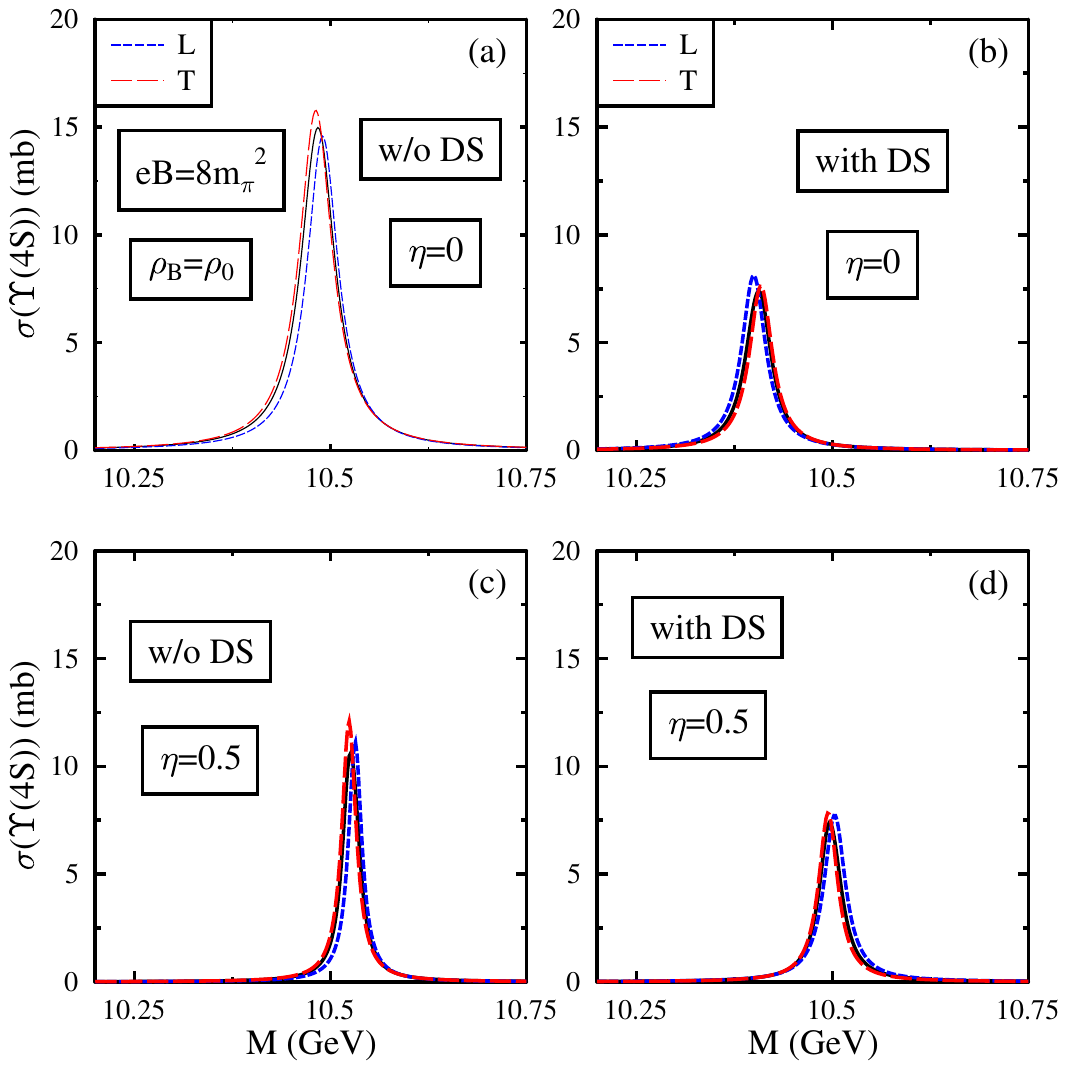}\hfill
\vskip -0.3in
    \caption{Same as figure 
\ref{Cross_section_Upsilon4s_rhb0_3mpi2},
for $eB=8 m_\pi^2$.}
    \label{Cross_section_Upsilon4s_rhb0_8mpi2}
\end{figure}
\begin{figure}
\vskip -2.2in
    \includegraphics[width=0.9\textwidth]{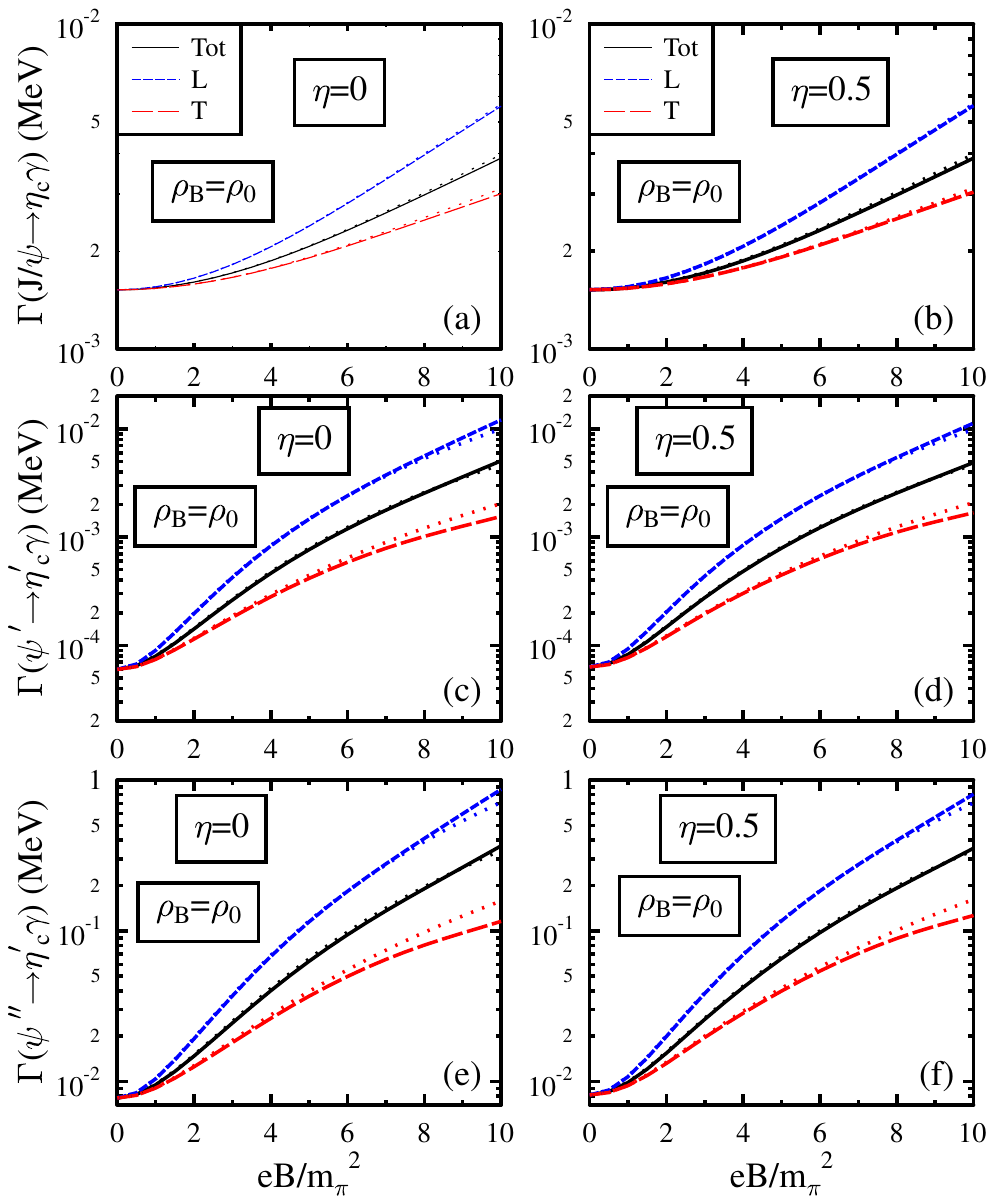}\hfill
\vskip -0.3in
    \caption{$\Gamma (V\rightarrow P\gamma)$ (in MeV)
for the charmonium states ($J/\psi \rightarrow \eta_c \gamma$, 
$\psi' \rightarrow \eta'_c \gamma$ 
and $\psi'' \rightarrow \eta'_c \gamma$), as functions of
$eB/{m_\pi}^2$, showing the contributions from the longitudinal
(L) and transverse (T) components, for symmetric and asymmetric nuclear
matter. 
These are compared with the case when the Dirac sea effects
are nor taken into account (shown as dotted lines). 
}
    \label{Charmonium_Rad_dw}
\end{figure}
\begin{figure}
\vskip -2.2in
    \includegraphics[width=0.9\textwidth]{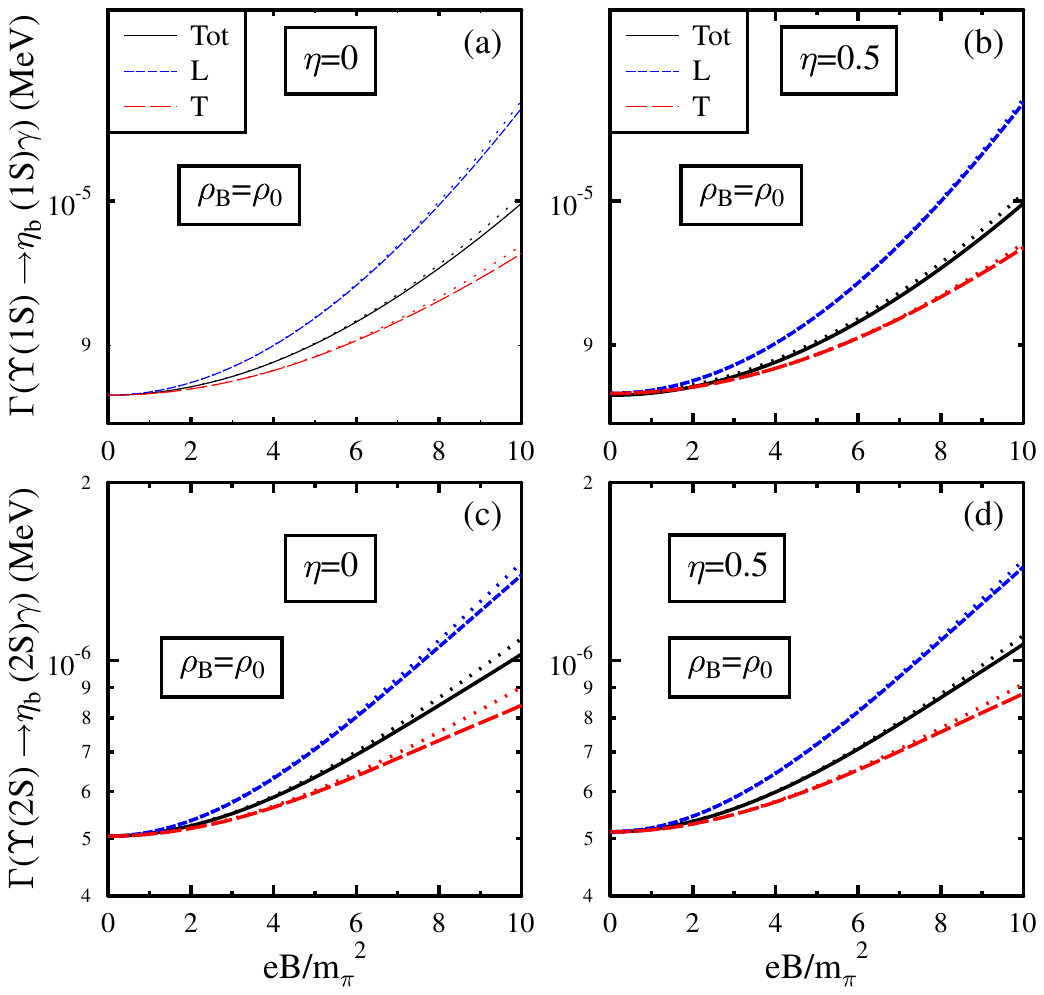}\hfill
\vskip -0.3in
    \caption{$\Gamma (V\rightarrow P\gamma)$ in MeV
    for the bottomonium states ($\Upsilon (1S) \rightarrow 
\eta_b (1S) \gamma$ and $\Upsilon (2S) 
\rightarrow \eta_b (2S) \gamma$), as functions of
$eB/{m_\pi^2}$ for symmetric and asymmetric nuclear
matter, showing the individual contributions from the longitudinal
(L) and transverse (T) components.
These are compared with the case when the Dirac sea effects
are nor taken into account (shown as dotted lines). }
   \label{Upsilon_Rad_dw_a}
\end{figure}
\begin{figure}
\vskip -2.2in
    \includegraphics[width=0.9\textwidth]{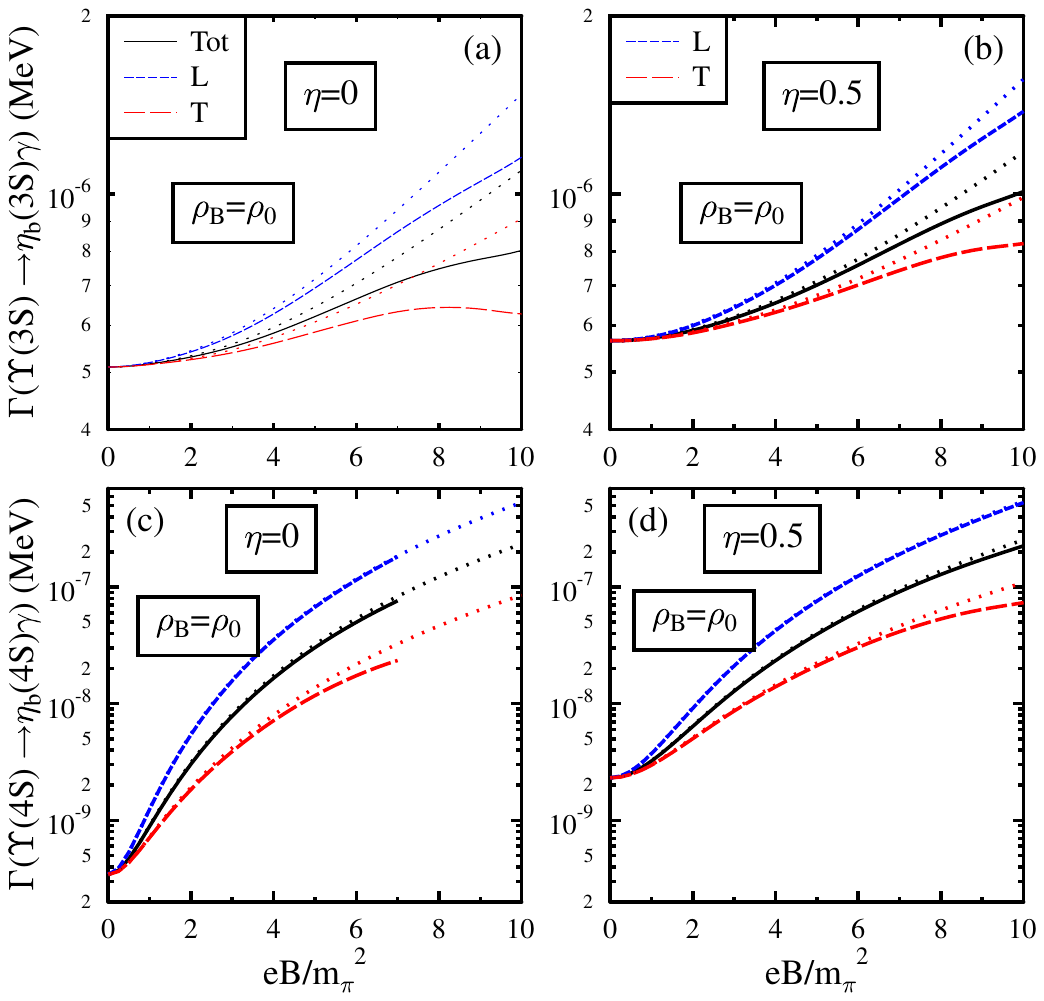}\hfill
\vskip -0.3in
    \caption{Same as figure \ref{Upsilon_Rad_dw_a}, for
$\Upsilon (3S) \rightarrow \eta_b (3S) \gamma$ and $\Upsilon (4S) 
\rightarrow \eta_b (4S) \gamma$.}
   \label{Upsilon_Rad_dw_b}
\end{figure}
The mass modifications due to PV mixing have been studied 
using an effective Lagrangian density of the form 
\cite{charmonium_mag_lee,Quarkonia_B_Iwasaki_Oka_Suzuki} 
\begin{equation}
{\cal L}_{PV\gamma}=\frac{g_{PV}}{m_{av}} e {\tilde F}_{\mu \nu}
(\partial ^\mu P) V^\nu,
\label{PVgamma}
\end{equation}
where $m_{\rm {av}}=(m_V+m_P)/2$, $m_P$ and $m_V$ are the masses 
of the pseudoscalar and vector mesons,
${\tilde F}_{\mu \nu}$ is the dual electromagnetic field.
In the presence of the PV mixing effects (incorporated
through the phenomenological Lagrangian (\ref{PVgamma})), 
the masses of the pseudoscalar and the longitudinal component
of the vector meson are given as \cite{charmonium_mag_lee}
\begin{equation}
m^{2\ {(PV)}}_{P,V^{||}}=\frac{1}{2} \Bigg ( M_+^2 
+\frac{c_{PV}^2}{m_{av}^2} \mp 
\sqrt {M_-^4+\frac{2c_{PV}^2 M_+^2}{m_{av}^2} 
+\frac{c_{PV}^4}{m_{av}^4}} \Bigg),
\label{mpv_long}
\end{equation}
where $M_\pm^2=m_V^2\pm m_P^2$ and $c_{PV}= g_{PV} eB$. 
The coupling parameter $g_{PV}$
is fitted from the observed value of the radiative decay width
for $V\rightarrow P\gamma$ for the charmonium sector. 
However, due to lack of radiative decay data 
for the bottomonium states, we estimate the modifications to the masses
of the bottomonium pseudoscalar and vector mesons \cite{upslndw_mag}
due to mixing of these states in the presence of a magnetic field,  
using the Hamiltonian 
\cite{Alford_Strickland_2013,Zhao_Prog_Part_Nucl_Phys_114_2020_103801,Quarkonia_B_Iwasaki_Oka_Suzuki}
\begin{equation}
H_{\rm {spin-mixing}}=-{\sum _{i=1}^2} 
{\vec \mu}_i \cdot {\vec B},
\label{H_spin_mixing}
\end{equation}
which decribes the interaction of the magnetic 
moments of the constituent quark (antiquark) with the external magnetic field.
In the above, 
${\vec \mu}_i =g|e|{q_i} {\vec {S_i}}/(2m_i)$ 
is the magnetic moment of the $i$-th particle, $g$ is the Lande g-factor
(taken to be $2(-2)$ for the quark(antiquark)), $q_i$, $\vec {S_i}$,
$m_i$ are the electric charge (in units of the magnitude of the
electronic charge, $|e|$), spin and mass of the $i$-th particle
\cite{charmonium_mag_lee,Quarkonia_B_Iwasaki_Oka_Suzuki}.
This interaction leads to the mass of the 
pseudoscalar (longitudinal component of the vector meson)
to be modified to \cite{Alford_Strickland_2013}
\begin{eqnarray}
{m^{(PV)}_{P(V^{||})}}_{\rm {spin-mixing}}&=&
m_{P(V)} \mp \frac{\Delta E}{2} 
\Big ( (1+\Delta ^2)^{1/2}-1\Big)\nonumber\\
&\equiv & m_{P(V)}\mp {\Delta M}^{PV},
\label{m_PV}
\end{eqnarray}
where, $\Delta=2g|eB|((q_1/m_1)-(q_2/m_2))/\Delta E$,
and, $\Delta E=m_V-m_P$ is the difference in the masses 
of the  vector and pseudoscalar mesons.
In equations (\ref{mpv_long}) and (\ref{m_PV}), 
$m_V$ and $m_P$ refer to the masses calculated
from the medium change of the dilaton field
within the chiral effective model.
The mass modifications of the heavy quarkonium states
as discussed above, solely arise due to the coupling
of the spin angular momenta of the constituent  
of the constituent heavy quark (antiquark) with the external
magnetic field, as decsribed by the Hamiltonian given by
equation (\ref{H_spin_mixing}). These lead to drop (increase)
in the mass of the pseudoscalar (longitudinal component
of the vector) heavy quarkonium state, whereas, the masses
of the transverse components of the vector meson remain
unaffected due to the PV mixing.

The PV mixing effects for the masses of the charmonium states 
\cite{charmdw_mag} and the open charm ($D$ and $\bar D$) mesons 
are incorporated using the phenomenological Lagrangian interaction
given by equation (\ref{PVgamma}) and for the bottomonium states
and open bottom mesons, due to unavailability of radiative decays
data, using the Hamiltonian interaction given by equation
(\ref{H_spin_mixing}) for the study of the heavy quarkonia
decay widths. There is observed to be 
appreciable mass modifications of the pseudoscalar and the longitudinal
component of the vector charmonium states,
due to the PV mixings \cite{charmdw_mag}. 
In the present work, the effects of the Dirac sea contributions 
are also considered on the masses of the 
of the heavy quarkonium as well as open heavy flavour mesons
and their effects on the partial decay widths to open charm (bottom)
mesons using a field theoretic model of composte hadrons.

\subsection{Decay widths of Heavy Quarkonia}

The partial decay widths of the heavy quarkonia 
to open heavy flavour mesons
in the magnetized nuclear matter are calculated from
the in-medium masses of the initial and final state
particles. These have been studied using the $^3P_0$ model 
for the charmonium sector \cite{charmdecay_mag}, 
as well as, using a field theoretic 
model of composite hadrons \cite{amspmwg,amspm_upsilon} for both the
charmonium and bottomonium sectors \cite{charmdw_mag,upslndw_mag}.
The mass modifications of the charmonium states
and the open charm mesons are observed to lead
to substantial modification of the partial decay width 
of $\psi(1D)\rightarrow D\bar D$ \cite{charmdw_mag}
due to $\psi(1D)-\eta_c(2S)$ mixing, as well as, due to
$D(\bar D)-D^* (\bar D^*)$ mixing effects 
\cite{open_charm_mag_AM_SPM}. The in-medium masses of the
upsilon states and open bottom mesons, including the PV mixing
effects and their effects on the partial decay width of
$\Upsilon (4S)\rightarrow B\bar B$ have been studied in 
Ref. \cite{upslndw_mag}.

In this subsection, we briefly describe the field theoretical 
model of composite hadrons \cite{spm781,spm782,spmdiffscat},
used to study the partial decay widths of the vector 
heavy quarkonium states to open heavy flavour mesons 
in magnetized (nuclear) matter. As the matter produced
in peripheral ultra-relativistic heavy ion collision 
experiments (where strong magnetic fields are created)
is extremely dilute,
we study the decay widhts of the charmonium state $\psi(3770)$ 
and the bottomonium state $\Upsilon(4S)$, which are the 
lowest states which decay to $D\bar D$ and $B\bar B$ 
in vacuum. 
The model used for the calculation of the decay widths
describes the hadrons as comprising of 
quark (and antiquark) constituents. 
The constituent quark field operators of the hadron in motion
are constructed from the constituent quark field operators of
the hadron at rest, by a Lorentz boosting.
Similar to the MIT bag model \cite{MIT_bag}, 
where the quarks (antiquarks) occupy
specific energy levels inside the hadron, it is assumed 
in the present model for the composite hadrons that 
the quark (antiquark) constituents carry fractions
of the mass (energy) of the hadron at rest (in motion) 
\cite{spm781,spm782}.
With explicit constructions of the charmonium (bottomonium)
state and the open charm (bottom) mesons, the decay width
of the heavy quarkonium state to open heavy flavour mesons.
is calculated using the light quark antiquark pair creation
term of the free Dirac Hamiltonian 
for constituent quark field \cite{amspmwg}.
The relevant part of the quark pair creation term is through the
 $q\bar q (q=u,d)$ creation for decay 
of the charmonium (bottomonium) state, $\Psi$ ($\Upsilon$), 
to the final state, $D\bar D$ ($B\bar B$).  
The pair creation term is given as 
\begin{equation}
{\cal H}_{q^\dagger\tilde q}({\bf x},t=0)
=Q_{q}^{(p)}({\bf x})^\dagger (-i 
\mbox{\boldmath $\alpha$}\cdot
\mbox{\boldmath $\bigtriangledown$} 
+\beta M_q)
{\tilde Q}_q^{(p')}({\bf x}) 
\label{hint}
\end{equation}
where, $M_q$ is the constituent mass of the light quark (antiquark). 
The subscript $q$ of the field operators in equation (\ref{hint})  
refers to the fact that the light antiquark, $\bar q$ and 
light quark, $q$ are the constituents 
of the $D(B)$ and $\bar D(\bar B)$ mesons with momenta ${\bf p}$ and
${\bf p'}$ respectively in the final state of the
decay of the charmonium (bottomonium) state, 
$\Psi(3770)(\Upsilon(4S))$.
The decay width of the quarkonium state, $M$, 
for the decay process $M\rightarrow F\bar F$, 
with $(M,F,\bar F)\equiv(\psi(3770),D,\bar D),(\Upsilon(4S),B,\bar B$), 
is calculated from the matrix element of 
the light quark-antiquark pair creation part of the free Dirac Hamiltonian,
between the initial quarkonium state and the final state mesons
for the reaction
$M\rightarrow F ({\bf p}) \bar F ({\bf p'})$
as given by
\begin{eqnarray}
\langle F ({\bf p}) | \langle {\bar F} ({\bf p'})|
{\int {{\cal H}_{q^\dagger\tilde q}({\bf x},t=0)d{\bf x}}}
|M^m (\vec 0) \rangle 
= \delta({\bf p}+{\bf p}')A_{M} (|{\bf p}|)p_m,
\label{tfi}
\end{eqnarray}
where,
$A_M(|{\bf p}|)$ is a polynomial of the magnitude of the
momentum, $|{\bf p}|$ of the outgoing mesons.
The decay width is calculated to be
\begin{eqnarray}
\Gamma(M\rightarrow F({\bf p}) {\bar F} (-{\bf p}))
= \gamma_\psi^2\frac{8\pi^2}{3}|{\bf p}|^3
\frac {p^0_{F}(|{\bf p}|) p^0_{\bar F}(|{\bf p}|)}{m_{M}}
A_{M}(|{\bf p}|)^2
\label{gammapsiddbar}
\end{eqnarray}
with $p^0_{F ({\bar F})}(|{\bf p}|)
=\big(m_{F ({\bar F})}^2+|{\bf p}|^2\big)^{{1}/{2}}$, and, 
$|{\bf p}|$, the magnitude of the momentum of the outgoing 
$F(\bar F)$ meson is given as, 
\begin{equation}
|{\bf p}|=\Big (\frac{{m_M}^2}{4}-\frac {{m_F}^2+{m_{\bar F}}^2}{2}
+\frac {({m_F}^2-{m_{\bar F}}^2)^2}{4 {m_M}^2}\Big)^{1/2}.
\label{pd}
\end{equation}
In the above, the masses of the $F(\bar F)$ and heavy 
quarkonium state are the in-medium masses in the magnetized
nuclear matter calculated in the chiral effective model
including the effects of the Dirac sea of nucleons,
with additional contributions from lowest Landau levels
for the charged open charm (bottom) mesons, as well as,
effects due to PV mixing \cite{charmdw_mag_MC}.  
The parameter, $\gamma_M$, in the expression for 
the quarkonium decay width,
is a measure of the coupling strength
for the creation of the light quark antiquark pair,
to produce the $F\bar F$ final state. 
This parameter is adjusted to reproduce the
vacuum decay widths of $\psi(3770)$ to $D^+D^-$ and
$D^0 \bar {D^0}$ \cite{amspmwg} for the charm sector
and $\Upsilon(4S)\rightarrow
B^+B^-$ and  $\Upsilon(4S)\rightarrow
B^0 \bar {B^0}$ \cite{amspm_upsilon} for 
the bottom sector.

Including the PV mixing effect, the expression
for the decay width is modified to 
\begin{eqnarray}
\Gamma^{PV}(M &\rightarrow & F({\bf p}) {\bar F} (-{\bf p}))
=\gamma_M^2\frac{8\pi^2}{3}
\Bigg [ 
\Bigg(\frac{2}{3} |{\bf p}|^3 
\frac {p^0_F (|{\bf p}|) p^0_{\bar F}(|{\bf p}|)}{m_{M}}
A_{M}(|{\bf p}|)^2 \Bigg)
\nonumber \\
&+&\Bigg(\frac{1}{3} |{\bf p}|^3 
\frac {p^0_F(|{\bf p}|) p^0_{\bar F}(|{\bf p}|)}{m_{M}^{PV}}
A^{M}(|{\bf p}|)^2 \Bigg) \Big({|{\bf p}|\rightarrow |{\bf p|}
(m_{M}= m_{M}^{PV})}\Big)
\Bigg]. 
\label{gammapsiddbar_mix}
\end{eqnarray}
In the above, the first term corresponds to the transverse
polarizations for the quarkonium state, $M$, 
whose masses remain unaffected by the mixing of the 
pseudoscalar and vector quarkonium states. 
The second term in (\ref{gammapsiddbar_mix})
corresponds to the longitudinal component,
whose mass is modified 
due to the mixing with the pseudoscalar meson 
in the presence of the magnetic field.

\section{Production Cross-sections of Heavy Quarkonia}

The relativistic Breit Wigner spectral function
and the production corss-section (from scattering of particles
$a$ and $b$) of the vector meson $V$ are given as
\cite{Elena_16,Elena_17,Elena_13_1,BW_CS_Haglin,BW_CS_Li}
\begin{equation}
A_V (M)=C \cdot \frac{2}{\pi} 
\frac {M^2 \Gamma_V^*}{(M^2-{m_V^*}^2)^2+(M\Gamma_V^*)^2}
\label{Spectral_V}
\end{equation}
and
\begin{equation}
\sigma (M)=\frac{6 \pi^2
\Gamma_V^* A_V (M)}{q(m_V^*,m_a^*,m_b^*)^2}.
\label{sigma1}
\end{equation}
where $M$ is the invariant mass, $m_V^*$ and $\Gamma_V^*$
are the in-medium mass and decay width of the vector meson,
$m_a^*$ and $m_b^*$ are the in-medium masses of the scattered
particles.
$C$ in the expression for the spectral function
given by equation (\ref{Spectral_V}) is a constant 
determined by normalization condition
$\int _0 ^{\infty} A_V (M) d M=1.$
In equation (\ref{sigma1}), $q(m_V^*,m_a^*,m_b^*)$
is the momentum of the scattering particle $a(b)$ in the 
center of mass frame of the vector meson, $V$,
and is given as,
\begin{eqnarray}
q(m^*_V,m_a^*,m_b^*)
= \frac{1}{2M}
\Big (\big [{m_V^*}^2-(m_a^*+m_b^*)^2 \big]
\cdot \big [{m_V^*}^2-(m_a^*-m_b^*)^2 \big] \Big )^{1/2}.
\label{cm_mom}
\end{eqnarray}
The vector meson, $V$ may, however, be created from 
scattering of (as well as decay to) particles in different modes,
say particles $a_i$ and $b_i$ with masses $m^*_{a_i}$ 
and $m^*_{b_i}$ in the channel $i$. The in-medium decay width 
of the vector meson is then given as the sum of the decay 
widths in these channels, i.e., $\Gamma_V^*={\sum_i} {\Gamma_V^i}^* $.
The production cross-section of the vector meson, accounting for
all these channels is given as 
\begin{equation}
\sigma (M)=6 \pi^2 
\sum _i \frac{{\Gamma_V^i}^*}{q({m^{*}_V},m^*_{a_i},m^*_{b_i})^2}
A_V (M),
\label{sigmaV}
\end{equation}
where $q(m^*_V,m^*_{a_i},m^*_{b_i})$ is the
center of mass momentum of the particle $a_i$ as well as $b_i$ 
corresponding to the channel $i$ in the center of mass frame
of the vector meson, $V$. 

In the present work, we study the production cross-sections 
of the vector charmonium (bottomonium) states, 
$\psi(3770)$ and $\Upsilon(4S)$, which are the lowest states
which decay to open charm ($D\bar D$) and open bottom ($B\bar B$) 
mesons. The charmonium vector meson ($V\equiv \psi(3770)$)
is produced through scattering of $D^+D^-$ and $D^0 \bar {D^0}$
mesons,
and, 
the bottomonium state ($V\equiv \Upsilon (4S))$)
is produced through scattering of 
$B^+B^-$ and $B^0 \bar {B^0}$ mesons.
The masses of the heavy quarkonium states and the open
charm ($D$ and $\bar D$) and open bottom ($B$ and $\bar B$)
in the magnetized asymmetric nuclear matter
are calculated within the chiral effective model 
accounting for the nucleon Dirac sea and the anomalous magnetic
moments of the nucleons. The PV mixing is taken into account
to calculate the masses of the vector quarkonium states,
$\psi(3770)$ and $\Upsilon (4S)$, due to $\psi(3770)-\eta_c(2S)$
and $\Upsilon (4S)-\eta_b(4S)$ mixings, as well as,
for the open charm  $D (\bar D$) and open bottom mesons, 
$B(\bar B)$ due to $D-D^* (\bar D-\bar {D^*})$ and 
$B-B^* (\bar B-\bar {B^*})$ mixings. 
Including the PV effect introduces mass difference  
between the longitudinal and transverse components
of the vector state ($\psi(3770)$ and $\Upsilon(4S)$),
since the longitudinal component undergoes mass modification
due to mixing with the pseudoscalar meson ($\eta_c(2S)$ and
$\eta_b(4S)$), whereas the transverse component is
unaffected due to PV mixing.

In the presence of PV mixing, the production corss-section
of the vector meson, V, given by equation (\ref{sigma1}),
can be written in terms of the contributions from the
longitudinal and transverse components as
\begin{equation}
\sigma (M)=6 \pi^2 \Bigg (\frac{1}{3}
\sum _i \frac{{\Gamma_V^{*i}}^{L}}{q({{m^*_V}^L},m^*_{a_i},m^*_{b_i})^2}
 A_V^L (M) +\frac{2}{3}
\sum _i \frac{{\Gamma_V^{*i}}^{T}}{q({{m^*_V}^T},m^*_{a_i},m^*_{b_i})^2}
 A_V^T (M) \Bigg),
\label{sigmaV_PV}
\end{equation}
where, ${m^*_V}^{T(L)}$ and ${\Gamma_V^{*i}}^{T(L)}$ are
the mass and the decay width (in channel $i$) of the transverse 
(longitudinal) component of the vector meson, 
 $A_V^{T(L)} (M)$ is the contribution from the transverse (longitudinal)
component to the spectral function, given as,
\begin{equation}
A_V^{T(L)} (M)=C \cdot \frac{2}{\pi} 
\frac {M^2 \Gamma_V^{*T(L)}}{\Big(M^2-{{m^{*T(L)}_V}^2\Big)}^2
+\Big(M\Gamma_V^{*T(L)}\Big)^2},
\label{Spectral_V_TL}
\end{equation}
with the constant $C$ determined from the normalization condition
\begin{equation}
\int _0 ^{\infty} \Big(\frac{1}{3} A_V^L (M)+ \frac{2}{3} A_V^T (M) 
\Big) d M=1.
\end{equation}
 
\section{Radiative Decay widths of Heavy Quarkonia:}

The radiative decay widths of the vector meson, V
are obtained from the decay width as calculated 
using the interaction Lagrangian given by equation (\ref{PVgamma}).
The parameter, $g_{PV}$ is calculated from
the observed decay width of $V\rightarrow P\gamma$
in vacuum, for the charmonium states. 
However, due to lack of experimental data of radiative decays
for the upsilon states, we compute the
value of $g_{PV}$ for these states
using the expressions of the masses of the pseudoscalar meson
and the longitudinal component of the vector meson
with PV mixing, obtained from the phenomenological Lagrangian given
by (\ref{PVgamma}), in the limit of weak magnetic field.
The expressions for the masses (given by equation (\ref{mpv_long})), 
retaining terms upto the second order in $eB$ and leading order 
in $\frac{(m_V-m_P)}{(m_V+m_P)}$,
reduce to \cite{charmonium_mag_lee}
\begin{equation}
m^{2\ {(PV)}}_{P,V^{||}}=m^2_{P,V}
\mp \frac{{(g_{PV}eB)}^2}{(m_V^2-m_P^2)^2},
\label{mpv_long_weakB}
\end{equation}
where, $m_V$ and $m_P$ are the masses of the 
pseudoscalar and vector mesons as calculated within
the chiral effective model.
The values obtained for the parameter $g_{PV}$
for the processes $\Upsilon (1S)\rightarrow \eta_b(1S)\gamma$,
$\Upsilon (2S)\rightarrow \eta_b(2S)\gamma$,
$\Upsilon (3S)\rightarrow \eta_b(3S)\gamma$, and,
$\Upsilon (4S)\rightarrow \eta_b(4S)\gamma$, 
are 1.1726, 1.2436, 1.2850 and 1.2932, 
using which the values of radiative decay widths (in eV) 
in vacuum are obtained to be 8.70907, 0.4748802, 0.56039
and 0.0078332 respectively.

The radiative decay widths for the vector heavy quarkonium 
state are calculated in the magnetized nuclear matter,
from the in-medium masses for the vector and pseudoscalar mesons. 
In the presence of the PV mixing, the mass of the longitudinal component 
of the vector meson is modified due to mixing with the
pseudoscalar meson,  whereas, the mass of the transverse component 
is not affected by PV mixing. This leads to decays
of these components to be different, in the presence
of a magnetic field. 

The radiative decay width, 
$\Gamma(V\rightarrow P +\gamma)$ given as
\begin{eqnarray}
\Gamma (V\rightarrow P \gamma)
&=&\frac{e^2}{12}\frac{g_{PV}^2}{\pi m_{\rm {av}}^2}
\Bigg ({\frac{1}{3}{p^L_{cm}}^3 +\frac{2}{3}{p^T_{cm}}^3}
\Bigg)
\equiv  \frac{1}{3} \Gamma^L +\frac{2}{3} \Gamma^T 
\label{decay_VP}
\end{eqnarray}
where,
\begin{equation}
p^{L}_{\rm {cm}}
=\frac{{m^{(PV)}_{V^{||}}}^2-{m^{(PV)}_P}^2}{2m^{(PV)}_{V^{||}}}\,\,\,
{\rm and} \;\; p^{T}_{\rm {cm}}=\frac{m_V^2-{m^{(PV)}_P}^2}{2m_V},
\label{com_LT_Rad_DW}
\end{equation}
are the magnitudes of the center of mass momentum 
in the final state corresponding to the longitudinal 
and transverse components of the decaying vector meson. 

\section{Results and Discussions}
In the present study, the effects of the Dirac sea and 
PV mixing are investigated on the production cross-sections 
and radiative decays of the heavy quarkonia in magnetized 
nuclear matter. The production cross-sections are computed
from the masses and partial decay widths (to open charm (bottom)
mesons) of these states and the radiative decay widths,
$\Gamma (V\rightarrow P\gamma)$ are calculated 
from the in-medium masses of the vector ($V$) and
pseudoscalar ($P$) heavy quarkonium states in the magnetized matter.
The study of the effects of magnetic field can be important 
for observables of peripheral ultra-relativistic heavy ion collision 
experiments, e.g., at RHIC, BNL and LHC, CERN, 
where the produced magnetic fields are estimated to be 
extremely strong. The created matter is extremely dilute 
in these ultra-relativistic heavy ion collision experiments.
In the present work, we study the production cross-sections of 
$\psi(3770)$ and $\Upsilon (4S)$, which are the lowest 
states, which decay to $D\bar D$ and $B\bar B$
respectively, in vacuum.

The masses of the heavy quarkonium states 
in magnetized nuclear matter have been studied 
within a chiral effective model,
including the effects of the nucleon Dirac sea
in Ref. \cite{Heavy_Quarkonia_masses_MC}.
These are obtained from the medium modification
of the scalar dilaton field, $\chi$, which simulates
the gluon condensates of QCD.
For given values of the baryon density, $\rho_B$,
the isospin asymmetry parameter,
$\eta=(\rho_n-\rho_p)/(2\rho_B)$ (with $\rho_n$ and $\rho_p$
as the neutron and proton number densities), and, the 
magnetic field, $B$ (chosen to be along z-direction),
the value of the dilaton field is calculated through
the solution of the coupled equations
of motion of the dilaton field, $\chi$ and 
the scalar-isoscalar non-strange ($\sigma$),
the scalar-isoscalar strange ($\zeta$), and,
scalar-isovector non-strange ($\delta$) fields. 
The anomalous magnetic moments (AMMs) of the nucleons
are taken into consideration in the present study.
Due to presence of the external magnetic field,
there is mixing of the pseusoscalar (P) meson 
and the longitudinal component of the vector (V) meson
(PV mixing) which further modifies the masses of 
these mesons. The masses of the charmonium states
($J/\psi$, $\psi'\equiv \psi(2S)$, $\psi''\equiv \psi(1D)$,
$\eta_c$ and $\eta_c'\equiv \eta_c(2S)$), modified due to
PV mixing are calculated using the equation (\ref{mpv_long}),
obtained from the phenomenological Lagrangian density
given by equation (\ref{PVgamma}).
The parameter, $g_{PV}$ is calculated from
the observed decay width of $V\rightarrow P\gamma$
in vacuum. For the decay processes 
$J\psi\rightarrow \eta_c \gamma$,  
$\psi'\rightarrow \eta'_c \gamma$, 
and, $\psi^{''}\rightarrow \eta'_c \gamma$, the values
of $g_{PV}$ obtained are 2.094, 3.184 and 7.657
\cite{charmdw_mag}, consistent with the
vacuum decay widths (in keV) of 92.9, 0.2058 and 24.48 respectively
\cite{pdg}.
Due to lack of data on radiative decay widths for the bottomonium
states ($\Upsilon (NS)$ and $\eta_b(NS)$, for $N=1,2,3,4$), 
the masses of these mesons, as modified due to PV mixing
($\Upsilon(NS)-\eta_b(NS)$, $i=1,2,3,4$),
are calculated using the equation (\ref{m_PV}), obtained from 
the interaction Hamiltonian (\ref{H_spin_mixing}).
The masses of the charmonium and bottomonium states
have been calculated including the effects of Dirac sea
and PV mixing in Ref. \cite{Heavy_Quarkonia_masses_MC}.
The in-medium decay widths of $\psi(3770)\rightarrow D\bar D$
and $\Upsilon (4S)\rightarrow B\bar B$ have been studied
including the effects from the nucleon Dirac sea in Ref.
\cite{charmdw_mag_MC}. These have been computed from the mass
modifications of the decaying and produced mesons, 
using a field theoretical model of composite hadrons,
as described in section 2.2.
The PV mixing effects are also taken into account
for the study of mass modifications of the open heavy
flavour mesons, $D$, $\bar D$, $B$ and $\bar B$,
arising from $D-D^*$, $\bar D-{\bar {D^*}}$,
$B-B^*$ and $\bar B-{\bar {B^*}}$ mixings for the study
of the partial decay widths of the heavy quarkonium states
\cite{charmdw_mag_MC}. 

The production cross-sections of the states $\psi(3770)$
and $\Upsilon(4S)$ are computed from the in-medium masses
\cite{Heavy_Quarkonia_masses_MC},
and, the partial decay widths of these states to 
$D\bar D$ and $B\bar B$ respectively \cite{charmdw_mag_MC}. 
The DS and PV mixing effects are taken into account for the calculation
of the masses of the decaying and the produced mesons
for computing these partial decay widths. For the sake of
completeness, we show the masses and decay widths of the
heavy quarkonia 
\cite{Heavy_Quarkonia_masses_MC,charmdw_mag_MC}, 
which are used to calculate their production cross-sections.
In fig. \ref{Mass_DW_3770_rhb0}, the effects of the Dirac
sea as well as PV mixing on the mass, and, the partial decay width of 
$\psi(3770)$ to $D\bar D$, are shown for both the symmetric and
asymmetric nuclear matter for $\rho_B=\rho_0$ \cite{charmdw_mag_MC}. 
In the absence of the DS and PV effects, the in-medium mass of the 
charmoniun state, $\psi(3770)$ in the magnetized nuclear matter
arises due to the the medium modification of the scalar dilaton field, 
$\chi$ \cite{charmonium_mag}, which mimics the scalar gluon condensates 
of QCD within the chiral effective model.
As has already been mentioned, within the mean field approximation,
the dilaton field, $\chi$ and  the scalar fields,
$\sigma$, $\zeta$ and $\delta$ are solved from their equations of motion,
accounting for Landau level contributions for the proton.
These are solved for given values of the baryon density, $\rho_B$,
isospin asymmetry parameter, $\eta$ and the magnetic field, $\vec B
=(0,0,B)$. In Refs. \cite{dmeson_mag,charmonium_mag}, the effects 
of the anomalous magnetic moments of the nucleons 
and the isospin asymmetry of the magnetized nuclear
matter on the scalar fields were studied and observed 
to be appreciable only for high densities
and high values of the magnetic field. 
For $\rho_B=\rho_0$, both for symmetric and asymmetric
nuclear matter, the medium modifications of the scalar fields 
are marginal and hence the mass of the charmonium
is observed to be insensitive to the variation of the magnetic 
field \cite{charmonium_mag}, as can be seen from figure
\ref{Mass_DW_3770_rhb0}. 
The PV mixing effects on the mass of
$\psi(3770)$ (due to mixing of the longitudinal
component of $\psi(3770)$ with $\eta_c(2S)$), are observed
to be large, in the presence as well as absence of 
the Dirac sea effects. When the PV mixing effect is not taken into account,
there is observed to be a monotonic drop in the mass
of $\psi(3770)$ with increase in the magnetic field with
the inclusion of Dirac sea effects,
whereas, the mass remains almost unchanged in the absence of
nucleon Dirac sea effect. The PV mixing introduces a mass difference
between the longitudinal and transverse components of
$\psi(3770)$. The $\psi(3770)-\eta'_c$ mixing leads 
to an increase in the mass of the longitudinal component 
of $\psi(3770)$, whereas the transverse component
is unaltered due to the PV mixing.
The decay width of $\psi(3770)\rightarrow D\bar D$,
which is obtained after averaging over the contributions
from the longitudinal and transverse components,
is observed to rise appreciably with increase in the magnetic 
field, both for symmetric and asymmetric
nuclear matter \cite{charmdw_mag_MC}. 
The PV mixing leading to difference in masses
of the longitudinal and transverse components of the
vector state $\psi(3770)$ is observed as a distinct 
double peak structure in the production cross-sections 
of $\psi(3770)$ plotted in figures \ref{Cross_section_psi3770_rhb0_3mpi2}
and \ref{Cross_section_psi3770_rhb0_8mpi2} 
for $eB=3m_\pi^2$ and $eB=8m_\pi^2$  respectively.
The separation between the peaks is observed to be 
larger for the higher value of the magnetic field, 
for which the PV mixing is more appreciable.

In figure \ref{Mass_DW_Upsilon4S_rhb0}, the mass and decay width
(to open bottom mesons) of $\Upsilon (4S)$ are plotted as functions
of $eB/{m_\pi^2}$ at $\rho_B=\rho_0$ for symmetric and asymmetric 
nuclear matter, accounting for the nucleon Dirac sea 
as well as PV mixing effects \cite{charmdw_mag_MC}. 
In the absence of the DS and PV effects, similar to $\psi(3770)$
mass (shown in figure \ref{Mass_DW_3770_rhb0}), the mass of $\Upsilon (4S)$
is also observed to be insensitive to the variation of the magnetic field.
This is a reflection of the fact that for $\rho_B=\rho_0$, 
the change in the dilaton field, $\chi$ is negligible with variation 
in the magnetic field \cite{charmonium_mag}.  
The effects of PV mixing are observed to be much smaller 
on the mass of $\Upsilon (4S)$,
as compared to the effects from the Dirac sea.
This is contrary to the modification
of the mass of the charmonium state $\psi(3770)$, where both the PV mixing 
as well as the Dirac sea effects have significant contributions
to the mass of $\psi(3770)$. There is observed to be
a monotonic drop in the mass of $\Upsilon(4S)$ with the increase
in the magnetic field, when the PV mixing is not taken into 
account, but the Dirac sea contributions to the nucleon self energy
are taken into consideration. The mass shift is observed to be
much smaller for the isospin asymmetric nuclear matter, 
as compared to symmetric nuclear matter. 
In the presence of the Dirac sea contributions, 
the PV mixing is observed to lead to a smaller drop as compared 
to when it is ignored for $\eta$=0.5 case.  However, for symmetric nuclear
matter, there is observed to be a positive contribution
due to PV mixing upto $eB\sim 7 m_\pi^2$, above which the
contribution turns out to be negative. 
This is because the effective mass of the $\Upsilon (4S)$
turns out to be smaller than that of $\eta_b(4S)$,
as calculated within the chiral effective model,
which makes $\Delta E$ (and hence $\Delta M^{PV}$)
of equation (\ref{m_PV}) to be negative.
As might be observed from figure \ref{Mass_DW_Upsilon4S_rhb0},
There is observed to be an initial increase in the decay width 
of $\Upsilon(4S)$ with magnetic field
upto $eB\sim 3 m_\pi^2$, followed by a drop with further increase in
the magnetic field upto around $eB\sim 7 m_\pi^2$. At still higher
values of the magnetic field, there is seen to be again a drop.   
The decay width as calculated from a field theoretical model
of composite hadrons, has the dependence on the center 
of mass momentum, $|{\bf p}|$ (given by equation (\ref{pd})),
which depends on the in-medium masses of the decaying and
produced mesons, as a polynomial term multiplied by an exponential part,
which leads to the observed dependence of the decay width 
on the magnetic field. The dotted lines show the decay width
when the Dirac sea effects are not taken into account.
The values of the mass as well as the decay width are 
observed to be quite close for $eB=3m_\pi^2$ for both without
and with the Dirac sea effects, which is observed
as similar behaviour for the production cross-section
of $\Upsilon (4S)$, as shown in figure 
\ref{Cross_section_Upsilon4s_rhb0_3mpi2}.
For $\eta$=0, the mass of $\Upsilon (4S)$ is observed 
to have much larger drop as compared to the $\eta$=0.5 case
for $eB=8 m_\pi^2$, which is seen as a higher
downward shift of the peak position in the production
cross-section of $\Upsilon(4S)$ in
figure \ref{Cross_section_Upsilon4s_rhb0_8mpi2}.

The production cross-section of $\psi(3770)$
is plotted in figures \ref{Cross_section_psi3770_rhb0_3mpi2}
and  \ref{Cross_section_psi3770_rhb0_8mpi2} for 
$eB=3m_\pi^2$ and $eB=8m_\pi^2$ respectively.
There is observed to be a double peak structure
of the production cross-section of $\psi(3770)$,
which is is observed to be significantly more pronounced 
for the higher value of the magnetic field. 
This is due to the difference in the masses of the longitudinal and
transverse components of the vector meson (V) due to 
PV mixing, for the $\psi(3770)-\eta'_c$ mixing.
The production cross-section of $\Upsilon(4S)$
is plotted in figures \ref{Cross_section_Upsilon4s_rhb0_3mpi2}
and  \ref{Cross_section_Upsilon4s_rhb0_8mpi2} for 
$eB=3m_\pi^2$ and $eB=8m_\pi^2$ respectively.
However, since the PV mixing is much smaller compared
to the Dirac sea effect for the bottomonium sector,
the difference in the longitudinal and transverse contributions
to the production cross-sections for $\Upsilon(4S)$ are observed
to be marginal. There is observed to be a much larger 
(downward) shift of the peak position due to Dirac sea
contributions, for the higher magnetic field of 
$eB=8 m_\pi^2$ for the symmetric as compared to asymmetric
nuclear matter, as can be seen in
figure \ref{Cross_section_Upsilon4s_rhb0_8mpi2}.
The maximun value of the cross-section
is observed to drop when the Dirac sea effects are considered  
for the higher magnetic field of $eB=8 m_\pi^2$ and the effect
due to PV mixing, leading to different contributions
from the longitudinal and transverse components,
is still observed to be very small.

The PV mixing leads to different contributions from 
the longitudinal and transverse components (L and T) 
for the radiative deay widths ($V\rightarrow P \gamma$) 
for the processes. In figure \ref{Charmonium_Rad_dw},
the radiative decay widths of the charmonium states
($J/\psi \rightarrow \eta_c \gamma$, $\psi' \rightarrow \eta'_c \gamma$, 
and $\psi'' \rightarrow \eta'_c \gamma$) and in figures
\ref{Upsilon_Rad_dw_a} and \ref{Upsilon_Rad_dw_b}, 
for the bottomonium states (for the processes
$\Upsilon(NS)\rightarrow \eta_b(4S) \gamma$, $N=1,2,3,4$),
are plotted as functions of $eB/{m_\pi}^2$, showing the contributions 
from the longitudinal (L) and transverse components.
The radiative decay widths of the heavy quarkonium
states in the magnetized nuclear matter
are calculated from the mass modifications of 
the vector and pseudoscalar states and
the total radiative decay width is obtained by averaging 
over the contributions from the transverse and
longitudinal components using equation (\ref{decay_VP}).
The dependence of the radiative decay widths
on the masses of the vector and pseudoscalar mesons 
are through the center of mass momenta, $p_{\rm {cm}}^L$ and 
$p_{\rm {cm}}^T$ for the longitudinal
and transverse components of the vector meson, 
which are different due to the difference in the masses 
of these components arising from the PV mixing. 
When the DS effects are not taken into account,
the masses of the heavy quarkonium states for both the symmetric 
and asymmetric nuclear matter have negligible dependence
on the variation of the magnetic field in the absence 
of PV mixing effect. This leads to the mass of the longitudinal
component of the vector meson (calculated using equation 
(\ref{mpv_long})) also to be very similar in behaviour 
with change in the magnetic field, for the symmetric and asymmetric
nuclear matter, as can be seen from figures 
\ref{Mass_DW_3770_rhb0} and \ref{Mass_DW_Upsilon4S_rhb0}.
These lead to the radiative decay widths also to have 
similar dependence on the magnetic field
for symmetric as well as asymmetric nuclear matter,
both for the longitudinal and transverse components, 
hence for the total width which is calculated by averaging 
over the decay widths of these components, by using 
equation (\ref{decay_VP}).  
In the presence of the Dirac sea contributions, however,
significant drops in the masses of $\psi(3770)$ 
as well as of $\Upsilon (4S)$ with increase in the
magnetic field can be observed from figures \ref{Mass_DW_3770_rhb0}
and \ref{Mass_DW_Upsilon4S_rhb0}, when the PV mixing is not taken 
into account. The PV mixing leads to a significant 
increase in the mass of the longitudinal component 
of $\psi(3770)$, whereas, for $\Upsilon (4S)$,
the rise is observed to be rather moderate.
For $\rho_B=\rho_0$, the values of the masses (in MeV) of the 
psuodoscalar meson $\eta_c'$ (longitudinal component of $\psi(3770)$)
are observed to be 3548.4 (3638.5) at zero magnetic field
and 3185.1 (3624.2) for $eB=10 m_\pi^2$ in symmetric nuclear matter
and 3551.4 (3643) at $eB=0$ and 3247.7 (3680.5) for $eB=10 m_\pi^2$
in asymmetric nuclear matter (with $\eta$=0.5).
For the transverse
component of $\psi(3770)$, which is unaffected by PV mixing,
the mass (in MeV) is 3638.5 (3402.8) for $\eta$=0
and 3643 (3474.8) for $\eta$=0.5 at $eB=0 (10 m_\pi^2)$. 
The value of $p_{\rm {cm}}^L
(p_{\rm {cm}}^T)$ (in MeV) is modified from 88.9 for zero magnetic field
to 412.5 (210.7) for $eB=10 m_\pi^2$ in symmetric ($\eta$=0) nuclear matter
and from 90.4 at $B$=0 to 407.35 (219.63) for $eB=10 m_\pi^2$ 
in asymmetric ($\eta$=0.5) nuclear matter. 
For the same values of the magnetic fields,
the values of the decay widths (in MeV) for the longitudinal
and transverse components, which are proportional 
to ${p_{\rm {cm}}^L}^3$ and ${p_{\rm {cm}}^T}^3$ respectively, 
(as can be seen from equation (\ref{decay_VP})), 
are modified from 7.7725$\times 10^{-3}$ for
$eB=0$ to and 0.86736 (0.115636) at $eB=10 m_\pi^2$ for symmetric nuclear
matter and from 8.15$\times 10^{-3}$ for zero magnetic field
to 0.80654 (0.1264) at $eB=10 m_\pi^2$ for asymmetric ($\eta$=0.5) 
nuclear matter. As can be seen from figures 
\ref{Upsilon_Rad_dw_a} and \ref{Upsilon_Rad_dw_b},
the radiative decay widths of $\Upsilon(NS)\rightarrow 
\eta_b(NS)\gamma$, $N=1,2,3,4$ have similar qualitative 
dependence on the magnetic field for the symmetric and 
asymmetric nuclear matter. 
The modifications due to the DS effect is observed to be 
moderate for the radiative decay widths of the charmonium
and bottomonium states investigated in the present work,
except for $\Upsilon (3S)\rightarrow \eta_b(3S)\gamma$,
where the effect due to the nucleon Dirac sea contributions
is observed to be appreciable at higher values of the magnetic
field, both for the symmetric as well as asymmetric nuclear
matter. 
In the presence of the Dirac sea contributions,
for $\eta=0$, it is observed that the decay width 
$\Upsilon (4S)\rightarrow \eta_b(4S) \gamma$ is not possible,
for $eB$ larger than $7 m_\pi^2$, as might be seen 
from figure \ref{Upsilon_Rad_dw_b}. This is because
the mass of $\eta_b(4S)$ (modified due to PV mixing) 
becomes larger than the mass of both the transverse
(without PV mixing) and longitudinal component
(modified due to the PV mixng) $\Upsilon (4S)$,
because of which the center of mass momenta of radiative decays
(given by equaiton (\ref {com_LT_Rad_DW}))
are no longer positive definite.
There is observed to be increase in the radiative decay widths
in both the charm and bottom sectors.
The production cross-sections are observed to be modified 
appreciably for the vector charmonium state, $\psi(3770)$,
due to Dirac sea (DS) as well as PV mixing effects.
There is observed to be distinct double peak structure 
in the invariant mass spectrum of production cross-section
of $\psi(3770)$ due to the PV mixing, which might be observed 
in dilepton spectra. For the bottom sector, the DS effect is observed
to be the dominant magnetic field effect.
The observed enhancement of the radiative decay widths 
at high magnetic fields can affect the production of the vector 
and pseudoscalar heavy quarkonium states and the modifications
of the production corss-sections of the states $\psi(3770)$
and $\Upsilon(4S)$ can affect the yields of the heavy quarkonium
states and the open heavy flavour mesons 
in peripheral ultra-relativistic 
heavy ion collision experiments, where the magnetic fields
created can be huge.

\section{Summary}

To summarize, we have studied the production cross-sections
and radiative decay widths of the heavy quarkonium
(charmonium and bottomonium) states in magnetized nuclear matter. 
The effects of Dirac sea (DS) of nucleons as well as
the PV mixing (mixing of the pseudoscalar meson and the longitudinal
component of the vector meson) are taken into consideration 
for obtaining the mass modifications of the heavy quarkonium states
as well as open heavy flavour mesons within a chiral
effective model. The anomalous magnetic moments (AMMs)
of the nucleons are considered in the present study. 
The production cross-sections of $\psi(3770)$ and
$\Upsilon (4S)$ are obtained from the 
masses of these mesons, as well as, the partial 
decay width of $\psi(3770)$($\Upsilon(4S)$)
to $D\bar D$($B\bar B$), the latter computed using 
a field theoretical model of composite hadrons.
Contrary to the charm sector, where both PV mixing and the
DS contributions are observed to be important,
it is seen that the PV mixing has 
much smaller contribution as compared to the Dirac sea
contributions for the bottomonium states.
The PV mixing in the charm sector ($\psi(3770)-\eta_c(2S)$
mixing), which modifies the mass of only the longitudinal component
of the vector charmonium state $\psi(3770)$, is observed 
to be quite significant leading to appreciable difference 
in the masses of the longitudinal and the transverse components 
of the state ($\psi(3770)$).
This is seen as a double peak in the invariant mass
spectrum of the production cross-section of
$\psi(3770)$, which is observed to be much more pronounced
as the magnetic field is increased. 
If the produced magnetic field is strong enough,
this might be observed in the dilepton spectra as distinct peaks 
corresponding to the longitudinal and transverse components of the state
$\psi(3770)$. For the bottom sector, in the presence of the magnetic field,
the Dirac sea (DS) effect is the more dominant effect 
as compared to the PV mixing effect.
The radiative decay widths of the vector (V) heavy 
quarkonium state to pseudoscalar (P) are calculated
from the medium modifications of the masses of these mesons. 
There is observed to be appreciable increase in the 
values of the radiative decay widths in both the 
charm and bottom sectors. 
The modifications of the production cross-sections
and the radiative decay widths of the heavy quarkonia
in the magnetized matter should have observable consequences
on the production of heavy quarkonium states 
and open heavy flavour mesons, as these are created 
at the early stage of the non-central ultra-relativistic 
heavy ion collision experiments, when the magnetic field
can be still be extremely large.

\begin{section}*{Acknowledgements}
Amruta Mishra acknowledges financial support from Department 
of Science and Technology (DST), 
Government of India (project no. CRG/2018/002226) and
Ankit Kumar from University Grants Commission (UGC), 
Government of India (Ref. 1279/(CSIR-UGC NET June 2017)).
\end{section}


\begin{thebibliography}{10}
\bibitem{Hosaka}
A. Hosaka, T. Hyodo, K. Sudoh, Y. Yamaguchi, S. Yasui, Prog. Part. Nucl. Phys. \textbf{96}, 88 (2017).

\bibitem{tuchin}
K. Tuchin, Adv. High Energy Phys. \textbf{2013}, 490495 (2013).

\bibitem{Z_Wang}
Z. Wang, J. Jhao, C. Greiner, Z. Xu and P. Zhuang,
Phys. Rev. C {\bf 105}, L041901 (2022).

\bibitem{Igor_Particles_2022_5} 
I. A. Shovkovy, Particles 2022, 5, 442.


\bibitem{pes1} M.E. Peskin, Nucl. Phys. {\bf B156}, 365 (1979).
\bibitem{pes2}
G. Bhanot and M.E. Peskin, Nucl. Phys. {\bf B156}, 391 (1979).
\bibitem{voloshin}
M.B.Voloshin, Nucl. Phys. B154 ,365 (1979).
\bibitem{leeko} Su Houng Lee and Che Ming Ko,
Phys. Rev. C {\bf 67}, 038202 (2003).

\bibitem{kimlee}
Sugsik Kim, Su Houng Lee, Nucl. Phys. A {\bf 679}, 517 (2001).

\bibitem{klingl} F. Klingl, S. Kim, S. H. Lee, P. Morath and
W. Weise, Phys. Rev. Lett. {\bf 82}, 3396 (1999).

\bibitem{amarvjpsi_qsr}
Arvind Kumar and Amruta Mishra, Phys. Rev. C {\bf 82}, 045207 (2010).
\bibitem{jpsi_etac_mag}
Pallabi Parui, Ankit Kumar, Sourodeep De, Amruta Mishra,
arXiv: 1811.04622 (nucl-th).
\bibitem{moritalee}
K. Morita and S.H. Lee, Phys. Rev. C {\bf 77}, 064904 (2008);
S.H. Lee and K. Morita, Phys. Rev. D {\bf 79}, 011501(R) (2009);
K. Morita and S.H. Lee, Phys. Rev. C {\bf 85}, 044917 (2012);
K. Morita and S.H. Lee, Phys. Rev. Lett {\bf 100}, 022301 (2008).

\bibitem{open_heavy_flavour_qsr}
Arata Hayashigaki , Phys. Lett. B {\bf 487}, 96 (2000);
T. Hilger, R. Thomas and B. K\"ampfer, Phys. Rev. C {bf 79},
025202 (2009);
 T. Hilger, B. K\"ampfer and S. Leupold,
Phys. Rev. C {\bf 84}, 045202 (2011);
S. Zschocke, T. Hilger and B. K\"ampfer,
Eur. Phys. J. A {\bf 47} 151 (2011).

\bibitem{Wang_heavy_mesons}
Z-G. Wang and Tao Huang, Phys. Rev. C {\bf 84}, 048201 (2011);
Z-G. Wang, Phys. Rev. C {\bf 92}, 065205 (2015).

\bibitem{arvind_heavy_mesons_QSR}
Rahul Chhabra and Arvind Kumar,
Eur. Phys. J A {\bf 53}, 105 (2017);
Rahul Chhabra and Arvind Kumar,
Eur. Phys. J C {\bf 77}, 726 (2017);
Arvind Kumar and Rahul Chhabra,
Phys. Rev. C {\bf 92}, 035208 (2015).



%
%

\bibitem{qmc}
P. A. M. Guichon, Phys. Lett. B {\bf 200}, 235 (1988);
\bibitem{open_heavy_flavour_qmc}
        K. Tsushima, D. H. Lu, A. W. Thomas, K. Saito, and R. H. Landau,
        Phys. Rev. C {\bf 59}, 2824 (1999);
        A. Sibirtsev,   K. Tsushima, and A. W. Thomas,
        Eur. Phys. J. A {\bf 6}, 351 (1999).
        K. Tsushima and F. C. Khanna, Phys. Lett. B {\bf 552}, 138
        (2003).

\bibitem{Schechter} J. Schechter, Phys. Rev. D {\bf 21}, 3393 (1980).
\bibitem{paper3}
        P. Papazoglou, D. Zschiesche, S. Schramm, J. Schaffner-Bielich,
        H. St\"ocker, and W. Greiner, Phys. Rev. C {\bf 59},  411  (1999).
\bibitem{kristof1}
        A. Mishra, K. Balazs, D. Zschiesche, S. Schramm,
        H. St\"ocker, and W. Greiner,
        Phys. Rev. C {\bf 69}, 024903 (2004).
\bibitem{amarvdmesonTprc}
Arvind Kumar and Amruta Mishra, Phys. Rev. C {\bf 81}, 065204
(2010).
\bibitem{amarvepja}
Arvind Kumar and Amruta Mishra, Eur. Phys. A {\bf 47}, 164
(2011).
\bibitem{AM_DP_upsilon}
Amruta Mishra and Divakar Pathak, Phys. Rev. C {\bf 90}, 025201
(2014).

\bibitem {amdmeson}
A. Mishra, E. L. Bratkovskaya, J. Schaffner-Bielich,
S.Schramm and H. St\"ocker, Phys. Rev. {\bf C 69}, 015202 (2004).
\bibitem{amarindamprc}
Amruta Mishra and Arindam Mazumdar, Phys. Rev. C {\bf 79},  024908 (2009).
\bibitem{DP_AM_Ds}
Divakar Pathak and Amruta Mishra, Adv. High Energy Phys. 2015,
697514 (2015).
\bibitem{DP_AM_bbar}
Divakar Pathak and Amruta Mishra, Phys. Rev. C {\bf 91}, 045206
(2015).
\bibitem{DP_AM_Bs}
Divakar Pathak and Amruta Mishra, Int. J. Mod. Phy. E {\bf 23},
1450073 (2014).

\bibitem{friman}
B. Friman, S. H. Lee and T. Song, Phys. Lett, B {\bf 548}, 153
(2002).

\bibitem{3p0_1}
A. Le Yaouanc, L. Oliver, O. Pene and  J. C. Raynal,
Phys. Rev. D {\bf 8}, 2223 (1973).
\bibitem{3p0_2}
A. Le Yaouanc, L. Oliver, O. Pene and  J. C. Raynal,
Phys. Rev. D {\bf 9}, 1415 (1974).
\bibitem{3p0_3}
A. Le Yaouanc, L. Oliver, O. Pene and  J. C. Raynal,
ibid, Phys. Rev. D {\bf 11}, 1272 (1975).
\bibitem{3p0_4}
T. Barnes, F. E. Close, P. R. Page and E. S. Swanson, Phys. Rev. D
{\bf 55}, 4157 (1997).

\bibitem{amspmwg}
Amruta Mishra, S. P. Misra and W. Greiner, Int. J. Mod. Phys.
E {\bf 24}, 155053 (2015).
\bibitem{amspm_upsilon}
Amruta Mishra and S. P. Misra, Phy. Rev. C {\bf 95}, 065206 (2017).

\bibitem{dmeson_mag}
Sushruth Reddy P, Amal Jahan CS, Nikhil Dhale, Amruta Mishra, J. Schaffner-Bielich, Phys. Rev. C \textbf{97}, 065208 (2018).

\bibitem{bmeson_mag}
Nikhil Dhale, Sushruth Reddy P, Amal Jahan CS, Amruta Mishra, Phys. Rev. C \textbf{98}, 015202 (2018).

\bibitem{charmonium_mag}
Amal Jahan CS, Nikhil Dhale, Sushruth Reddy P, Shivam Kesarwani, Amruta Mishra, Phys. Rev. C \textbf{98}, 065202 (2018).

\bibitem{charmdecay_mag}
 A. Mishra , A. Jahan CS , S. Kesarwani , H. Raval , S. Kumar, and J.
Meena , Eur. Phys. J. A 55,99 (2019).

\bibitem{charmonium_mag_QSR} S. Cho, K. Hattori, S. H. Lee, K. Morita
and S. Ozaki, Phys. Rev. Lett. {\bf 113}, 122301 (2014).

\bibitem{charmonium_mag_lee}
S. Cho, K. Hattori, S. H. Lee, K. Morita
and S. Ozaki, Phys. Rev. D {\bf 91}, 045025 (2015).

\bibitem{Suzuki_Lee_2017}
K. Suzuki and S. H. Lee, Phys. Rev. C {\bf 96}, 035203 (2017).

\bibitem{Alford_Strickland_2013}
J. Alford and M. Strickland, Phys. Rev. D {\bf 88}, 105017
(2013).

\bibitem{Zhao_Prog_Part_Nucl_Phys_114_2020_103801}
J. Zhao, K. Zhou, S. Chen, P. Zhuang, Prog. Nucl. Part. Phys. 
{\bf 114} (2020) 103801.

\bibitem{Quarkonia_B_Iwasaki_Oka_Suzuki}
S. Iwasaki, M. Oka, K. Suzuki, Eur. Phys. Jour. A {\bf 57}
(2021) 222.


\bibitem{Open_bottom_MC}
Sourodeep De, Pallabi Parui and Amruta Mishra, Int. Jour. Mod. Phys.
E {\bf 31}, 2250106 (2022).

\bibitem{Heavy_Quarkonia_masses_MC}
Ankit Kumar and Amruta Mishra, arXiv: 2208:14962 (hep-ph).

\bibitem{charmdecay_mag_MC}
Sourodeep De, Pallabi Parui and Amruta Mishra, Phys. Rev. C {\bf 107},
074003 (2023).

\bibitem{charmdw_mag}
 Amruta Mishra, S.P. Misra,
Phys. Rev. C {\bf 102}, 045204 (2020).

\bibitem{open_charm_mag_AM_SPM}
Amruta Mishra and S. P. Misra, Int. Jour. Mod. Phys. E {\bf 30},
2150064 (2021).

\bibitem{upslndw_mag}
 Amruta Mishra, S.P. Misra,
Int. Jour. Mod. Phys. E \textbf{31} 06, 2250060 (2022).

\bibitem{charmdw_mag_MC}
Amruta Mishra and S.P. Misra, Phys. Rev. D {\bf 107}, 074003
(2023).

\bibitem{Preis}
F. Preis, A. Rebhan, and A. Schmitt, Lect. Notes Phys. \textbf{871}, 51 (2013).

\bibitem{menezes}
D. P. Menezes, M. Benghi Pinto, S. S. Avancini, and C.
Providencia, Phys. Rev. C \textbf{80}, 065805 (2009); D.P. Menezes, M. Benghi Pinto, S. S. Avancini, A. P. Martinez,
and C. Providencia, Phys. Rev. C \textbf{79}, 035807 (2009).

\bibitem{arghya}
Arghya Mukherjee, Snigdha Ghosh, Mahatsab Mandal, Sourav Sarkar, and Pradip Roy, Phys. Rev. D \textbf{98}, 056024 (2018).

\bibitem{haber}
Alexander Haber, Florian Preis, and Andreas Schmitt, Phys. Rev. D \textbf{90}, 125036 (2014). 



\bibitem{pdg}
 P.A. Zyla et al. (Particle Data Group), Prog. Theor. Exp. Phys. \textbf{2020}, 083C01 (2020).

\bibitem{spm781} S. P. Misra, Phys. Rev. D {\bf 18}, 1661 (1978).
\bibitem{spm782} S. P. Misra, Phys. Rev. D {\bf 18}, 1673 (1978).
\bibitem{spmdiffscat} S. P. Misra and L. Maharana, Phys. Rev. D
{\bf 18}, 4103 (1978).

\bibitem{MIT_bag}
A. Chodos, R. L. Jaffe, K. Johnson and C. B. Thorn,
Phys. Rev. D {\bf 10}, 2599 (1974).

\bibitem{spmddbar80} S.P.Misra, K. Biswal and B. K. Parida,
Phys. Rev. {\bf D 21}, 2029 (1980).

\bibitem{Elena_16}
A. Ilner, D. Cabrera, C. Markert and E. Bratkovskaya,
Phys. Rev. C {\bf 95}, 014903 (2017).

\bibitem{Elena_17}
A. Ilner, J. Blair, D. Cabrera, C. Markert and E. Bratkovskaya,
Phys. Rev. C {\bf 99}, 024914 (2019).

\bibitem{Elena_13_1}
A. Ilner, D. Cabrera, P. Srisawad and E. Bratkovskaya,
Nucl. Phys. A {\bf 927}, 249 (2014).

\bibitem{BW_CS_Haglin}
K. Haglin, Nucl. Phys. A {\bf 584}, 719 (1995).

\bibitem{BW_CS_Li}
G. Q. Li, C. M . Ko and G. E. Brown, Nucl. Phys. A {\bf 611},
539 (1996).

\end{thebibliography}
\end{document}